\documentclass[showpacs,aps,floatfix,superscriptaddress]{revtex4-1}
\usepackage{amsmath, amsthm, amssymb}
\usepackage{CJK}
\usepackage{graphicx}
\usepackage[justification=justified]{caption}
\usepackage[export]{adjustbox}
\usepackage{epsfig}
\usepackage{wasysym}
\usepackage{multirow}
\usepackage{dsfont}
\usepackage{graphicx}
\usepackage{caption}
\usepackage{subcaption}
\usepackage{float}
\usepackage{url}
\usepackage{hyperref}
\usepackage{bm}
\usepackage{verbatim}
\usepackage{booktabs}

\newcommand{\bra}[1]{\ensuremath{\langle \,#1\, \vert}}				                           
\newcommand{\ket}[1]{\ensuremath{\vert \,#1\, \rangle}}				                           
\newcommand{\braket}[2]{\ensuremath{\langle\, #1\, \vert \,#2\, \rangle} }

\usepackage{letltxmacro}
\makeatletter
\let\oldr@@t\r@@t
\def\r@@t#1#2{%
\setbox0=\hbox{$\oldr@@t#1{#2\,}$}\dimen0=\ht0
\advance\dimen0-0.2\ht0
\setbox2=\hbox{\vrule height\ht0 depth -\dimen0}%
{\box0\lower0.4pt\box2}}
\LetLtxMacro{\oldsqrt}{\sqrt}
\renewcommand*{\sqrt}[2][\ ]{\oldsqrt[#1]{#2}}
\makeatother

\makeatletter
\newcommand{\thefigurename}{Figure}
\def\fnum@figure{\thefigurename\ \thefigure}
\makeatother
\renewcommand{\thefigurename}{FIG.}

\newcommand\hksqrt[2][]{\mathpalette\DHLhksqrtA{{#1}{#2}}}
 \def\DHLhksqrtA#1#2{\setbox0=\hbox{$#1\DHLhksqrtB#2$}\dimen0=\ht0
 \advance\dimen0-0.3\ht0
 \setbox2=\hbox{\vrule height\ht0 depth -\dimen0}%
 {\box0\lower0.4pt\box2}}
 \def\DHLhksqrtB#1#2{\def\a{#1}\def\b{}\ifx\a\b\sqrt{#2\,}\else\sqrt[#1]{#2\,}\fi}

\begin{document}
\title{Improving Mean-Field Theory
for Bosons in Optical Lattices via
Degenerate Perturbation Theory}
\author{M. K{\"u}bler }
\thanks{martin\_kuebler@gmx.de}
\affiliation{Physics Dept.~and Res.~Center OPTIMAS, Technische Universit\"at
Kaiserslautern, 67663 Kaiserslautern, Germany}

\author{F. T. Sant'Ana}
\thanks{felipe.taha@usp.br}
\affiliation{S\~{ao} Carlos Institute of Physics, University of S\~{ao Paulo}, 13566-590, S\~{a}o Carlos, SP, Brazil}

\author{F. E. A. dos Santos}
\thanks{santos@ufscar.br}
\affiliation{Department of Physics, Federal University of S\~{a}o Carlos, 13565-905, S\~{a}o Carlos, SP, Brazil}

\author{A. Pelster}
\thanks{axel.pelster@physik.uni-kl.de}
\affiliation{Physics Dept.~and Res.~Center OPTIMAS, Technische Universit\"at
Kaiserslautern, 67663 Kaiserslautern, Germany}

\begin{abstract}
The objective of this paper is the theoretical description of the Mott-insulator to superfluid quantum phase transition of a Bose gas in an optical lattice. In former works the Rayleigh-Schr\"odinger perturbation theory was used within a mean-field approach, which yields partially non-physical results since the degeneracy between two adjacent Mott lobes is not taken into account. In order to correct such non-physical results we apply the Brillouin-Wigner perturbation theory to the mean-field approximation of the Bose-Hubbard model. Detailed explanations of how to use the Brillouin-Wigner theory are presented, including a graphical approach that allows to efficiently keep track of the respective analytic terms. To prove the validity of this computation, the results are compared with other works. Besides the analytic calculation of the phase boundary from Mott-insulator to superfluid phase, the condensate density is also determined by simultaneously solving two algebraic equations. The analytical and numerical results turn out to be physically meaningful and can cover a region of system parameters inaccessible until now. Our results are of particular interest provided an harmonic trap is added to the former calculations in an homogeneous system, in view of describing an experiment within the local density approximation. Thus, the paper represents an essential preparatory work for determining the experimentally observed wedding-cake structure of particle-density profile at both finite temperature and hopping.

\end{abstract}


\maketitle

\section{Introduction}

Since the first realization of a Bose-Einstein condensate in 1995 \cite{articleWiemannCornell, articleKetterle}, the field of ultracold quantum gases receives an ongoing strong interest to study a vast variety of new quantum many-body effects \cite{articleLegget, bookSmith, bookStringari, articleZwerger,booklewenstein}. Regarding optical lattices \cite{articleZoller}, one of these new effects is the quantum phase transition from a Mott-insulator to a superfluid phase \cite{articleHaensch}. This can be described theoretically via the Bose-Hubbard model \cite{articleBoseHubbard,articleHubbard}, which is a paradigm for quantum phase transitions \cite{bookSachdev}. There are many well-established methods to actually calculate the phase boundary of the Mott-insulator to superfluid phase transition. The purely analytic mean-field approach \cite{articleFisher}, which is as well used in this work, gives good qualitative insights about the physics close to the phase boundary, but it is quantitatively imprecise as a drawback. As a contrast, a full numerical Quantum Monte-Carlo simulation \cite{articleSvistunov} yields quantitatively quasi-exact results, but its qualitative insights are limited. In lower dimensions, a strong-coupling expansion \cite{articleFreericks} gives good results, while for higher dimensions an effective action approach \cite{articleSchuetzhold,Bradlyn,articleEdnilson,grass,grass2} is more reliable. Another method is the process chain, which allows to extend both the strong-coupling expansion \cite{wang} and the effective action approach \cite{A,B} to higher orders. Thus, it became possible to yield for the quantum phase boundary an accuracy comparable to Quantum Monte-Carlos simulations and even to determine critical exponents \cite{articleHolthaus}. Also, an effective action approach to handle a time-periodic driven optical lattice was studied in \cite{wang2}. In Ref.~\cite{cheng} it became even possible to reconstruct experimentally the homogeneous superfluid to Mott-insulator quantum phase transition for a two-dimensional ultracold quantum gas in an optical lattice with an additional harmonic confinement via an in-situ imaging.

This paper deals with the problem of determining the condensate density for a homogeneous Bose gas in an optical lattice within mean-field theory. As in the vicinity of the mean-field phase boundary the condensate density is supposedly small, the standard approach starts with the mean-field Hamiltonian \cite{articleFisher} and determines the ground-state-energy with non-degenerate perturbation theory \cite{thesisHoffmann}. However, the resulting Landau expansion \cite{articleLandau} yields a condensate density that turns out to vanish between two adjacent Mott lobes and has, therefore, to be considered as not enough accurate. The origin of this non-physical result stems from the fact that between adjacent Mott lobes a degeneracy occurs, so that in this point the non-degenerate perturbation theory is no longer valid. This deficiency was recognized, for instance, in Ref.~\cite{articleMelo} and solved tentatively by determining the condensate density with degenerate perturbation theory. Although this allowed to obtain a non-vanishing condensate density between two adjacent Mott lobes, the result is inconsistent insofar as the condensate density does not vanish at the mean-field phase boundary. Thus, the fundamental problem remained of how to combine the results from non-degenerate \cite{thesisHoffmann} and degenerate \cite{articleMelo} perturbation theory in order to obtain a consistent mean-field result for the condensate density.

The present paper solves this problem by using the Brillouin-Wigner perturbation theory \cite{bookBrillouinWigner}. It is based on a projection formalism, which allows to eliminate a larger fraction of the Hilbert space in order to obtain an effective eigenvalue equation for the remaining subspace. The resulting effective Hamiltonian can then be systematically expanded in a power series of the perturbative term. In this way, it turns out that the Brillouin-Wigner perturbation theory formally interpolates between the non-degenerate and the degenerate perturbation theory.


In the context of the Bose-Hubbard mean-field theory, we proceed as follows. Section \ref{sec2} introduces the state of the art for analytically describing the Mott insulator-superfluid quantum phase transition, pointing out what modern theories can do and where they fail. In the following Section \ref{sec3}, we overcome all these problems by applying the Brillouin-Wigner perturbation theory. This allows to determine reliably the quantum phase boundary and the condensate density in the superfluid phase. Finally, we consider, in Section \ref{Trap}, the effect of an additional harmonic trap to our calculations within the local density approximation, motivated by the experimental detection of the wedding cake structure that was reported in \cite{widera}. Our results allow to study the melting of the characteristic density profile in form of a wedding-cake structure due to the mutual impact of both thermal fluctuations and finite hopping. This leads, in particular, to the emergence of superfluid shells between the Mott lobes as has already been studied in Ref.~\cite{gerbier}.
\section{\label{sec2}The problem}

In this section we describe the current problem by calculating the condensate density. To this end, we first present the Bose-Hubbard model to describe bosons in an optical lattice, then we introduce within the Landau theory the condensate wave function as an order parameter to distinguish between the Mott and the superfluid phase. Afterwards, we apply the mean-field theory together with non-degenerate perturbation theory to get an approximate result for the quantum phase boundary. Hence, we get formulas for the phase boundary and the order parameter, where the latter turns out to be physically inconsistent.

\subsection{Bose-Hubbard model}\label{sec:Bose-Hubbard Hamiltonian}

The Bose-Hubbard model, first published in 1963 by H. A. Gersch and G. C. Knollman \cite{articleBoseHubbard}, is a bosonic adapted version of the Hubbard model, which was published by J. Hubbard earlier in 1963 \cite{articleHubbard} for fermionic particles. Two main assumptions are made for the Bose-Hubbard model. The first one is that the temperature is so low, that it is sufficient to take into account only the lowest energy band. The second assumption is to neglect any long-range interaction and long-range hopping.

The Hamilton operator for the Bose-Hubbard model reads
\begin{align}\label{eq.1.4}
\hat{H}=\frac{1}{2}U\sum_{i}\hat{n}_{i}\left( \hat{n}_{i}-1 \right) -J\sum_{\langle i,j \rangle}\hat{a}^{\dagger}_{i}\hat{a}_{j} -\mu \sum_{i}\hat{n}_{i}\,,
\end{align} 
with $U$ denoting the on-site interaction to be either $U > 0$ (repulsive) or $U < 0$ (attractive), whereas $\hat{a}^{\dagger}_i$ and $\hat{a}_i$ are the bosonic creation and annihilation operators at site $i$, while $\hat{n}_i=\hat{a}_i^\dagger \hat{a}_i$ denotes the number operator at site $i$. Furthermore, $J$ represents the single-particle Hamiltonian, also called the hopping term. The summation indices $\langle i,j \rangle$ represent the restriction that only nearest neighboring transitions are allowed. Finally, $\mu$ denotes the chemical potential, which corresponds within a grand-canonical description to the energy for adding a boson to the optical lattice.

\subsection{Landau theory}\label{sec:Landau Theory}

According to Landau \cite{articleLandau,bookLandauLifschitz}, we can represent the energy of our system as a polynomial function of the order parameter, i.e. $E\left( \Psi^*, \Psi \right)$. Because of the $U(1)$-symmetry present in the Bose-Hubbard Hamiltonian (\ref{eq.1.4}), this dependency reduces to $E\left( \Psi^{*}\Psi \right)$ and only even orders can be present in the expansion
\begin{align}\label{Landaureihe}
E=a_0 + a_2 \Psi^* \Psi+ a_4 \Psi^{*2} \Psi^2+... \,.
\end{align}

Following the Landau approach to describe second-order phase transitions, we seek to minimize the truncated energy where terms of order higher than four are neglected provided that $a_4 > 0$. With this we find the extrema by differentiation
\begin{align}
\frac{\partial E}{\partial \Psi^*}=\Psi \left( a_2+ 2 a_4 \Psi^* \Psi \right) \,.
\end{align}

With $\partial E / \partial \Psi^* =0$, this gives two possible solutions for the condensate density $\Psi^* \Psi$, either we have
\begin{align}\label{PP=0}
\Psi^* \Psi =0
\end{align}
or
\begin{align}\label{eq.OP...}
\Psi^* \Psi = - \frac{a_2}{2 a_4} \,.
\end{align}
Note that the minima of $E$ depend on the sign of $a_2$. For $a_2>0$ we have the Mott insulator phase where there is no condensate density, thus (\ref{PP=0}) describes such a phase. This determines the energy of the Mott-insulator according to (\ref{Landaureihe}) as
\begin{align}
E_{\rm Mott}=a_0\,.
\end{align}
On the other hand, for $a_2<0$ the minima of $E$ are given by (\ref{eq.OP...}). In order to obtain the energy in the superfluid phase we have to insert (\ref{eq.OP...}) into (\ref{Landaureihe}) and get
\begin{align}
{E}_{\rm Superfluid}=a_0 - \frac{a_2^2}{4a_4}\,.
\end{align}
In addition, the boundary separating the superfluid and the Mott-insulator phase is given by the points in the quantum phase diagram where $a_2=0$.

\subsection{Mean-field approximation}\label{sec:Mean-Field Approximation}

The energy $E$ can be calculated via a field-theoretic method, where the Legendre transform of the grand-canonical free energy gives very precise results \cite{Ednilson,articleEdnilson}. Another way is to apply the mean-field approximation, which is quantitatively less correct, but gives already a quite good qualitative insight. Furthermore, the calculations are less complex and thus much faster to perform with high precision.

Due to the non-local term present in the hopping term of (\ref{eq.1.4}) a direct calculation turns out to be difficult. In order to get rid of this non-local term approximatively, we perform a Bogoliubov decomposition,
\begin{align}\label{eq.1.10}
\hat{a}_{i}&=\Psi + \delta \hat{a}_{i}\, ,
\end{align}
with $\Psi$ representing the mean field, whereas $\delta \hat{a}_{i}$ stands for the fluctuation correction. Within the mean-field approximation one neglects all quadratic fluctuations, resulting in the Bose-Hubbard mean-field Hamiltonian,
\begin{align}\label{eq.MFHamil}
\hat{H}_{\rm{MF}}=& \frac{1}{2}U\sum_{i}\hat{n}_{i}\left( \hat{n}_{i}-1 \right)-\mu \sum_{i}\hat{n}_{i} 
 -Jz\sum_{i}\left( \Psi^{*}\hat{a}_{i}+\Psi\hat{a}_{i}^{\dagger}-\Psi^{*}\Psi \right) \,.
\end{align}
Here $z$ denotes the number of nearest neighbors. Since (\ref{eq.MFHamil}) is local, we can restrict ourselves effectively to one lattice site.

\subsubsection{Non-degenerate perturbation theory}
As the condensate density $\Psi^* \Psi$ is zero in the Mott-insulator and positive in the superfluid phase, we can assume that the order parameter is small as long as we stay in the superfluid phase close to the quantum phase boundary. This implies that corrections due to the kinetic term can be obtained in power series of $\Psi^*$ and $\Psi$ through a perturbative approach. In order to do so, we split the on-site mean-field Hamiltonian into an unperturbed part
\begin{equation}\label{H0}
	\hat{H}^{(0)}=\frac{1}{2}U\hat{n}\left( \hat{n}-1 \right) -\mu \hat{n}
\end{equation}
and a perturbation
\begin{equation}\label{V}
	\hat{V}=-Jz(\Psi^{*}\hat{a}+\Psi\hat{a}^{\dagger} -\Psi^{*}\Psi) \, ,
\end{equation}
with $\lambda$ denoting a smallness parameter according to
\begin{equation}\label{H}
	 \hat{H}=\hat{H}^{(0)}+\lambda \hat{V} \, .
\end{equation}

From standard non-degenerate perturbation theory we can get the energy in the Landau expansion up to the fourth order following Ref.~\cite[(3.39)]{thesisHoffmann}. Thus we have for the coefficients of (\ref{Landaureihe}):
\begin{align}\label{a0=E0}
a_0=E_{n}^{(0)}\,,
\end{align}
\begin{align}\label{eq.RSa2}
a_2=J z + J^2 z^2 \left( \frac{n+1}{E_{n}^{(0)}-E_{n+1}^{(0)}} + \frac{n}{E_{n}^{(0)}-E_{n-1}^{(0)}} \right)\,,
\end{align}
and
\begin{align}\label{eq.RSa4}
\notag
a_4=&J^4 z^4  \left [ \frac{n+1}{\left( E_{n}^{(0)}-E_{n+1}^{(0)} \right)^2} \left(
 \frac{n+2}{ E_{n}^{(0)}-E_{n+2}^{(0)} } \right. \right.
  \left. \left. -\frac{n}{ E_{n}^{(0)}-E_{n-1}^{(0)} } -\frac{n+1}{ E_{n}^{(0)}-E_{n+1}^{(0)} } \right) \right.
\\
& \left. + \frac{n}{\left( E_{n}^{(0)}-E_{n-1}^{(0)} \right)^2} \left(
\frac{n-1}{ E_{n}^{(0)}-E_{n-2}^{(0)} } \right. \right.
 \left. \left. -\frac{n+1}{ E_{n}^{(0)}-E_{n+1}^{(0)} }
 -\frac{n}{ E_{n}^{(0)}-E_{n-1}^{(0)} } \right) \right]\,.
\end{align}
Here the unperturbed ground-state energy is defined via 
\begin{equation}\label{EN0uncurly}
	E_n^{(0)} = \frac{1}{2} Un(n-1)-\mu n \, .
\end{equation}

According to Landau's theory, the phase boundary can be calculated from the condition $a_2=0$. The resulting equation is solved with respect to $J z / U$ as in Ref.~\cite{articleFisher}:
\begin{align}\label{QWERT}
\frac{Jz}{U}=\frac{- \left( E_{n}^{(0)} - E_{n+1}^{(0)} \right) \left( E_{n}^{(0)} - E_{n-1}^{(0)} \right)}{U\left[ E_{n}^{(0)} - E_{n-1}^{(0)} + 2 n E_{n}^{(0)} - n \left( E_{n+1}^{(0)} + E_{n-1}^{(0)} \right)\right]}\, .
\end{align}
\normalsize

For large $Jz/U$, we are in the superfluid phase, far away from the phase boundary, as the Mott-insulator needs low hopping probabilities. Since all of our theory is based on the assumption of being close to the quantum phase boundary, we cannot obtain reliable results for values of $Jz/U$ deep in the superfluid phase. Nevertheless, for $Jz/U\lesssim 0.35$, we assume our model to be valid. While for $Jz/U=0$, we have no superfluid phase and only a Mott insulator, we always reach the superfluid phase by increasing $Jz/U$. Another way to get from the Mott insulator to the superfluid phase is by tuning $\mu/U$ at $Jz/U>0$. If we start in the first Mott lobe and increase $\mu/U$, the ordered structure breaks down at some point and the superfluid phase is energetically more favorable and thus realized. For $\mu/U<0$, the system is in the superfluid phase for $Jz/U > - \mu/U$, whereas for $Jz/U < - \mu/U$ we have no particles at all.

\begin{figure}[t]
\centering
      \includegraphics[width=.6\columnwidth]{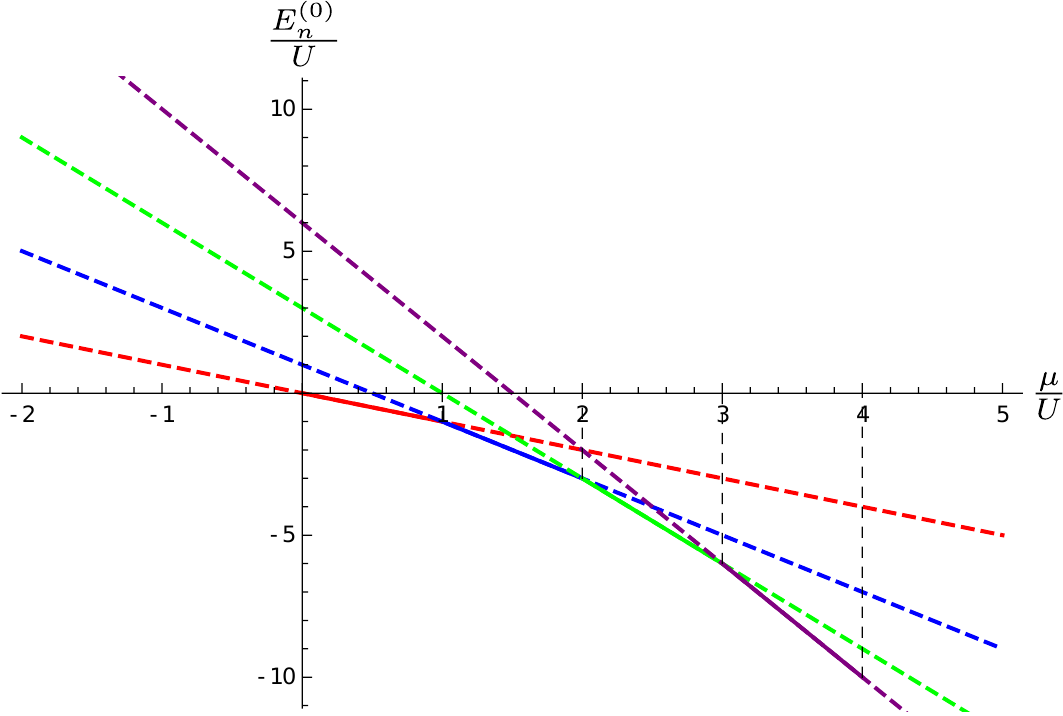}
\caption{(color online) Unperturbed ground-state energies (\ref{EN0uncurly}). Different lines correspond to different values for $n$ from smaller to larger slope: $n=1$ (red), $n=2$ (blue), $n=3$ (green), and $n=4$ (purple). Vertical dashed black lines correspond to the points of degeneracy. Solid colored lines represent realized lowest energy, while dashed colored lines indicate the continuation of the energy line.} \label{fig.Energies}
\end{figure}

After having obtained the quantum phase boundary, we take a closer look at the lowest energies for increasing $n$. In the plot of the unperturbed energies (\ref{EN0uncurly}) in FIG. \ref{fig.Energies}, we see that the ground state energies have a degeneracy at integer values of $\mu/U$. Like in between the lobes for $n=1$ (line with the smallest slope, red) and $n=2$ (line with the second smallest slope, blue) at $\mu/U=1$, we are at the degeneracy point of the energies $E_{1}^{(0)}$ and $E_{2}^{(0)}$. Analogous formulae are valid between every two neighboring lobes. It is exactly this degeneracy at $\mu = Un$ which makes every algebraic treatment of this system quite complex, but since we have always only two degenerate energies to handle at once, a solution can be found.

With this degeneracy in mind, we now discuss the order parameter. First, we plot (\ref{eq.OP...}) by using (\ref{eq.RSa2}) and (\ref{eq.RSa4}). Since $a_4$ approaches infinity for $\mu = U n$, where we have $E_{n}^{(0)}=E_{n+1}^{(0)}$, according to (\ref{a0=E0}), the condensate density $\Psi^* \Psi$ tends to zero at the degeneracy between two adjacent lobes, which falsely indicates a quantum phase boundary. This non-physical behavior is depicted in FIG. \ref{fig.OPRS} through the dashed (orange) plot.

\subsubsection{Degenerate perturbation theory}
One way to improve these results is to apply degenerate perturbation theory, which was done up to the first perturbative order in Ref.~\cite{articleMelo}. Since two degenerate states are taken into account, for further references, we name it the two-states approach, it results in a 2$\times$2-matrix
\begin{align}\label{DRSPTG1}
\Gamma^{(1)}=\begin{pmatrix}
E_{n}^{(0)} + Jz \lambda \Psi^{*} \Psi & -\lambda Jz \Psi^* \sqrt{n+1}  \\
-\lambda Jz \Psi \sqrt{n+1} & E^{(0)}_{n+1}+Jz \lambda \Psi^{*} \Psi
\end{pmatrix}\,,
\end{align}
where the matrix entries are calculated up to first order in $\lambda$. Inserting the explicit expressions for $E^{(0)}_{n}$ and $E^{(0)}_{n+1}$ from (\ref{EN0uncurly}) the eigenvalues of $\Gamma^{(1)}$ read
\begin{align}\label{eq.2.84}
E_{n\pm}=& \lambda Jz\Psi^* \Psi + \frac{1}{2} \left[U n^2 -2\mu(n+1)\right]
 \pm  \frac{1}{2}\hksqrt{\left( \mu - Un\right)^2 +4\lambda^2 J^2 z^2 \Psi ^* \Psi \left(n+1\right)}\,.
\end{align}
Now we extremize the energy (\ref{eq.2.84}) with respect to the condensate density $\Psi^* \Psi$ by applying $\partial E_{n\pm}/\left(\Psi \partial \Psi ^{*}\right)=0$, yielding
\begin{equation}\label{eq.2.92}
\Psi ^* \Psi =\frac{ \left(n+1\right)}{4} - \frac{\left( \mu - Un \right)^2}{4\lambda^2 J^2 z^2  \left(n+1\right)}\,,
\end{equation}
which coincides with \cite{articleMelo}. Note that both the cases with positive and negative sign yield the same condensate density.

\begin{figure}[t]
	\centering
\includegraphics[width=.6\textwidth]{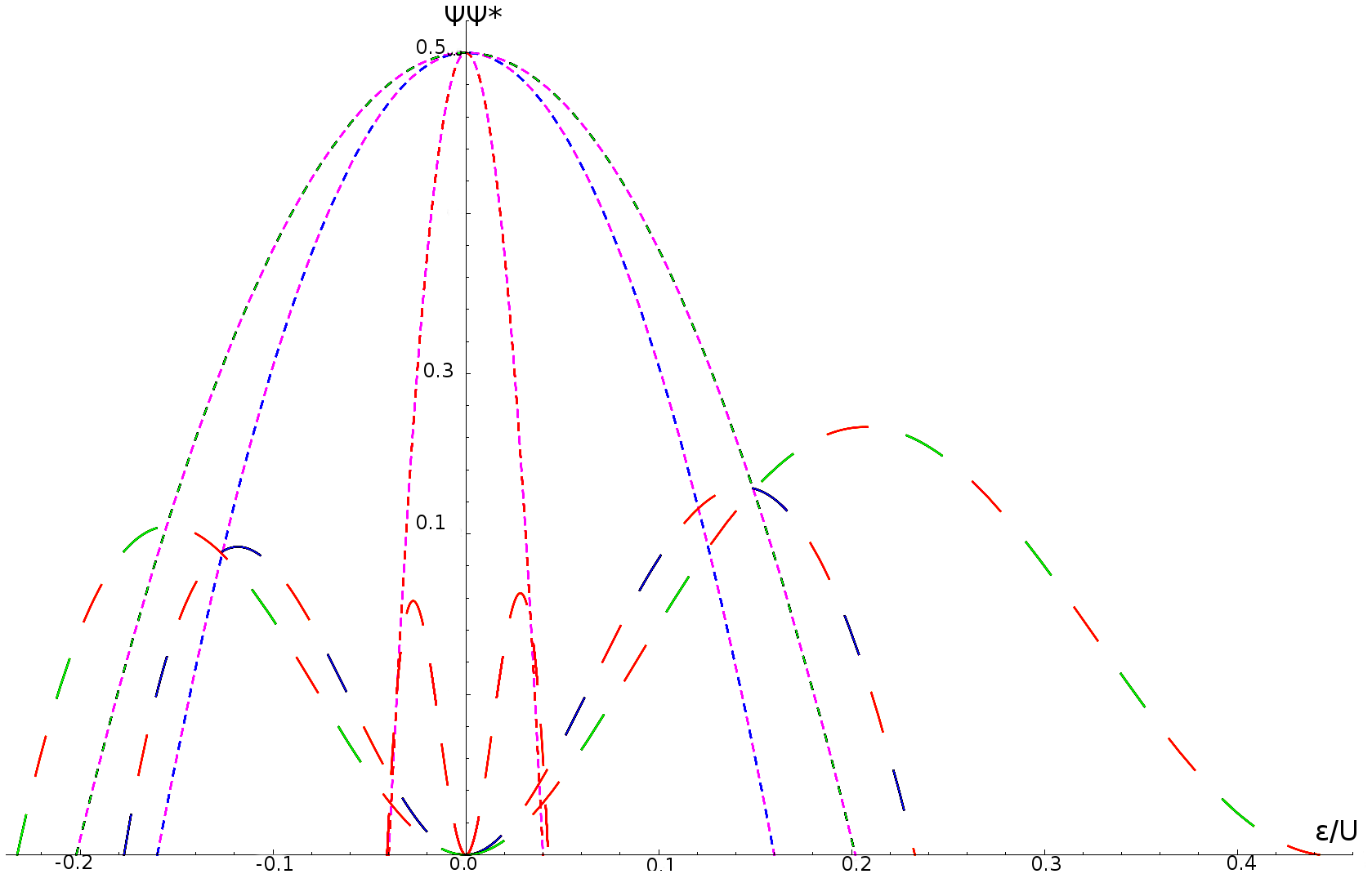}
\caption{(color online) Condensate density from non-degenerate perturbation theory according to (\ref{eq.OP...}) with (\ref{eq.RSa2}) and (\ref{eq.RSa4})\cite{thesisHoffmann} (orange, dashed) in comparison with the condensate density from degenerate perturbation theory according to (\ref{OPMelovarepsilon}) \cite{articleMelo} (magenta, dotted) with $\mu=Un+\varepsilon$ and $n=1$ for the left part and $n=2$ for the right part, respectively. From the spacing inside to the outside we have $Jz/U = 0.02$ (red), $Jz/U = 0.08$ (blue), and $Jz/U = 0.101$ (green). Dashed (orange) plots are zero at the mean-field quantum phase boundary, yielding an unphysical behavior at the degeneracy, having increasing maxima for increasing $Jz/U$, and for $Jz/U = 0.101$ and $\varepsilon/U = 0.442$ the lobe is just touching in one point and goes smoothly to zero. The dotted  (magenta) plots give a physical behavior at the degeneracy, but has always the value $\Psi^* \Psi =0.5$ at the degeneracy, which can directly be seen in (\ref{OPMelovarepsilon}). For small $Jz/U$ and close to quantum the phase boundary, the plots coincide.} \label{fig.OPRS}
\end{figure}

At the degeneracy we have $J=0$, which would lead to a quadratic divergent term in (\ref{eq.2.92}). But for the degeneracy $E^{(0)}_{n} =E^{(0)}_{n+1}$, we get $\mu - Un = 0$, which appears as well in the numerator. Thus we have no divergence problem here. Let us now introduce the parameter $\varepsilon$ according to $\mu=Un+\varepsilon$ in order to analyze the nearly-degenerate case. If $\varepsilon=0$, we are at the degeneracy, for positive and negative small $\varepsilon$, we are nearly degenerate and can describe the direct vicinity of the degeneracy following Ref.~\cite{articleMelo} according to
\begin{align}\label{OPMelovarepsilon}
\Psi ^* \Psi =\frac{ \left(n+1\right)}{4} - \frac{\varepsilon^2}{4\lambda^2 J^2 z^2  \left(n+1\right)}\,,
\end{align}
which is depicted in the dotted (magenta) plot of FIG. \ref{fig.OPRS}.

\begin{figure}[t]
	\centering
	\includegraphics[width=.6\columnwidth]{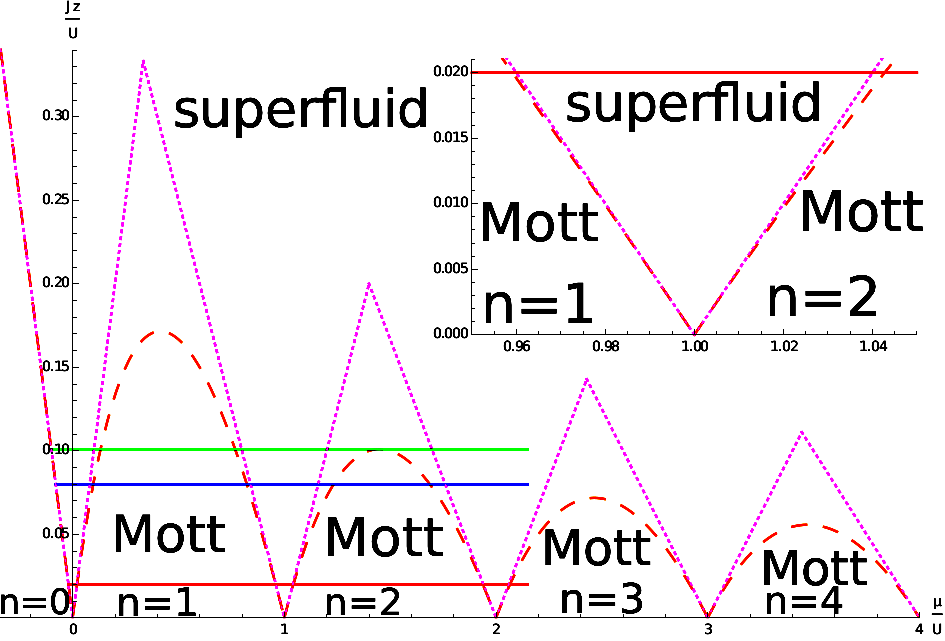}
	\caption{(color online) Quantum phase boundary, obtained by Rayleigh-Schr\"odinger perturbation theory. The non-degenerate theory \cite{thesisHoffmann} yields the dashed orange plot, while the degenerate theory \cite{articleMelo} reproduces the dotted magenta plot. Inside the lobes we are in the Mott-insulator phase, while outside the lobes we are in the superfluid phase. The number of particles $n$ increases from left to right by one per lobe. The three horizontal lines are from bottom to top $Jz/U=0.02$ (red), $Jz/U=0.08$ (blue), and $Jz/U=0.101$ (green). They all start at the line $Jz/U=-\mu/U$, which indicates $n=0$ and end at $\mu/U=2.15$. The inset shows the part between the first two Mott lobes with increased size, with the same axis as the big plot.}
	\label{fig.vieleLobes}
\end{figure}

By setting $\Psi^* \Psi =0$ in (\ref{eq.2.92}) we obtain the quantum phase boundary shown in the dotted (magenta) plot in FIG. \ref{fig.vieleLobes}. The quantum phase boundary obtained out of the degenerate approach is always linear, which is only coinciding with the non-degenerate case for $n=0$. Nevertheless, for small values of $Jz/U$, this linearization is a good approximation (see inset in FIG. \ref{fig.vieleLobes}). The tips of these triangular, dotted (magenta) Mott lobes are at $\mu/U=1/3 \approx 0.333$, $\mu/U=7/5=1.4$, $\mu/U=17/7\approx2.429$, and $\mu/U=31/9\approx 3.444$ for increasing $n$, which is not the same value as for the tips of the curved, dashed (orange) lobes, which are correspondingly at $\mu/U=\sqrt{2}-1 \approx 0.414$, $\mu/U=\sqrt{6}-1 \approx 1.449$, $\mu/U=2 \sqrt{3}-1 \approx 2.464$, and $\mu/U=2\sqrt{5}-1 \approx 3.472$. These values coincide more for higher $\mu/U$. The horizontal lines are from top to bottom at $Jz/U=0.02$ (red), $Jz/U=0.08$ (blue), and $Jz/U=5-2\sqrt{6} \approx 0.101$ (green), while the latter one hits the second lobe exactly on its tip. These lines allow a better comparison between the dashed (orange) and the dotted (magenta) quantum phase boundary.

\section{\label{sec3}The solution}

By comparing FIG. \ref{fig.vieleLobes} with FIG. \ref{fig.OPRS}, we conclude that the non-degenerate approach (dashed, orange) yields a reasonable quantum phase boundary, but an inconsistent condensate density, while the degenerate approach (dotted, magenta) yields an improved result for the order parameter, but a worse quantum phase boundary. Therefore, in order to handle both adequately, another approach is necessary. To this end, we stay in a perturbative picture, which already succeeded in reproducing the quantum phase boundary, but in order to get the order parameter as well we will apply the Brillouin-Wigner perturbation theory, which is summarized in Appendix \ref{sec:Appendix A: Derivation of Brillouin-Wigner}.

\subsection{One-state approach}\label{One-State Approach}

At first we tackle our problem within the one-state approach of the Brillouin-Wigner perturbation theory as specified in Section \ref{subonestate} of the Appendix. To this end we consider a subspace of the Hilbert space spanned by only one eigenstate $\ket{\Psi_{n}^{(0)}}$ and its projector operator
\begin{align}\label{eq.6.1}
\hat{P}=\ket{\Psi_{n}^{(0)}} \bra{\Psi_{n}^{(0)}}\,.
\end{align}
The ground-state energy is then identified with $E_{n}=\bra{\Psi_{n}^{(0)}}\hat{H}_{\rm eff}\ket{\Psi_{n}^{(0)}}$. From (\ref{eq.2.31,006}) up to third order in $\lambda$ and inserting $\hat{H}^{(0)}$ and $\hat{V}$ from (\ref{H0}) and (\ref{V}) yields
\begin{align}\label{eq.8.11}
\nonumber
E_{n}=&E_{n}^{(0)} + \lambda J z \Psi^{*}\Psi + \lambda ^2 J^2 z^2 \Psi^{*}\Psi \left(  \frac{n}{E_{n} - E_{n-1}^{(0)}}\right. \left. + \frac{n+1}{E_{n} - E_{n+1}^{(0)}}\right) \\
& +\lambda ^3 J^3 z^3 \left(\Psi^{*}\Psi\right)^2 \left[  \frac{n}{\left(E_{n} - E_{n-1}^{(0)}\right)^2} + \frac{n+1}{\left(E_{n} - E_{n+1}^{(0)}\right)^2}\right]
\,.
\end{align}
Note that (\ref{eq.8.11}) represents a self-consistency equation of the energy $E_n=E_n(\Psi^* \Psi)$.

\subsubsection{Quantum phase boundary}

The mean-field quantum phase boundary was already shown in FIG. \ref{fig.vieleLobes} (dashed orange line) obtained from the Rayleigh-Schr\"odinger perturbation theory. Here we will reproduce this result within the one-state approach from the Brillouin-Wigner perturbation theory. In order to get the phase boundary we evaluate $\partial E_n\left(\Psi^*\Psi\right)/\left(\Psi \partial\Psi^*\right)$, with $E_n$ being the energy formula from the one-state approach up to the third order in $\lambda$ according to (\ref{eq.8.11}).

We show now in a general way that we can neglect all terms with $\lambda$ of order 3 and higher. To this end we must observe the generic structure of $E_n\left(\Psi^*\Psi\right)$ in (\ref{eq.8.11}):
\begin{align}\label{eq.(1.18)}
E_n\left(\Psi^*\Psi\right) &=  \alpha + \Psi ^* \Psi \beta + \frac{\Psi ^* \Psi \gamma_0}{\gamma_1 + \Psi ^* \Psi \gamma_2} + \sum_{m\geq 2}^{\infty}\frac{ \left( \Psi ^* \Psi \right)^m k_m}{P\left( \Psi ^* \Psi \right)} \,.
\end{align}
The coefficients $\alpha$, $\beta$, $\gamma_0$, $\gamma_1$, $\gamma_2$, and $k_m$ are independent of $\Psi^* \Psi$, while $m$ is a natural number and $P(\Psi^* \Psi)$ is a polynomial. Performing the differentiation in (\ref{eq.(1.18)}), i.e.
\begin{align}\label{eq.25}
\frac{1}{\Psi}\frac{\partial E_n\left(\Psi^*\Psi\right)}{\partial\Psi^*}&= \beta +  \frac{\gamma_0 \gamma_1}{\left( \gamma_1 + \Psi ^* \Psi \gamma_2 \right)^2}+ \sum_{m\geq 2}^{\infty} \left( \frac{m\left( \Psi ^* \Psi \right)^{m-1}k_m P\left( \Psi ^* \Psi \right) }{P\left( \Psi ^* \Psi \right)^2} \right.
\left. - \frac{ \left( \Psi ^* \Psi \right)^{m}k_m \frac{1}{\Psi}\frac{\partial}{\partial \Psi ^*} P\left( \Psi ^* \Psi \right)}{P\left( \Psi ^* \Psi \right)^2} \right)\,,
\end{align}
we obtain for the quantum phase boundary
\begin{align}\label{eq.(1.19)}
\frac{1}{\Psi}\frac{\partial E_n \left(\Psi^*\Psi\right)}{\partial\Psi^*}\Bigg|_{\Psi ^* \Psi=0} = \beta + \frac{\gamma_0}{\gamma_1}\,.
\end{align}
Here we see that all corrections to higher order than 2 in $\lambda$ can be neglected. Thus, the phase boundary does not change even if higher orders in $\lambda$ are taken into account. 

\begin{figure}[t]
	\centering
\includegraphics[angle =270, width=.6\textwidth]{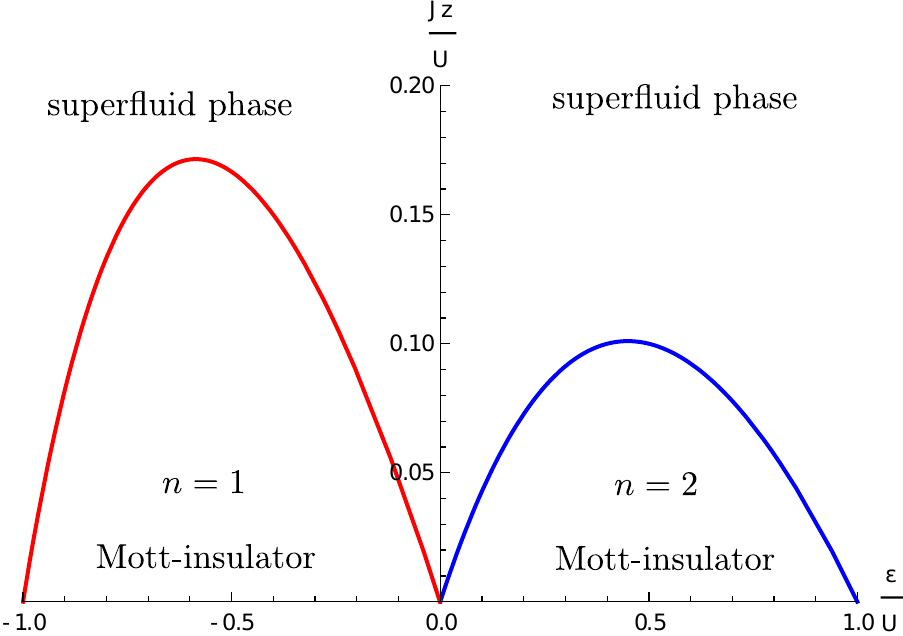}
\caption{(color online) Mott lobes representing the mean-field phase boundary with $\varepsilon = \mu/U -1$.} \label{fig.(1.99)}
\end{figure}

Comparing (\ref{eq.25}) with (\ref{eq.8.11}) we identify the relevant coefficients to be
\begin{gather}
\beta=\lambda J z\,,\\
\gamma_0=\lambda ^2 J^2  z^2 \left[(2n+1)E_{n}+(n-1)E_{n-1}^{(0)} -n E_{n+1}^{(0)} \right]\,,\\
\gamma_1=\left(E_{n}-E_{n+1}^{(0)}\right) \left(E_{n}-E_{n-1}^{(0)}\right)\,.
\end{gather}
Inserting them into (\ref{eq.(1.19)}) we obtain
\begin{gather}
\frac{1}{\Psi}\frac{\partial E_n \left(\Psi^*\Psi\right)}{\partial\Psi^*}\Bigg|_{\Psi ^* \Psi=0}  = \lambda J z + \lambda ^2 z^2 \frac{E_{n}-E_{n-1}^{(0)} + 2 n E_{n}-n E_{n+1}^{(0)} + n E_{n-1}^{(0)}}{\left(E_{n}-E_{n+1}^{(0)}\right) \left(E_{n}-E_{n-1}^{(0)}\right)}  \,.\label{eq29}
\end{gather}
Putting (\ref{eq29}) to zero we obtain
\begin{align}\label{form.6.7}
\frac{Jz}{U}=&-\frac{1}{\lambda  U}\frac{\left(E_n-E_{n+1}^{(0)}\right) \left(E_n-E_{n-1}^{(0)}\right)}{E_n-E_{n-1}^{(0)} + 2n E_n -nE_{n+1}^{(0)}-nE_{n-1}^{(0)}} \,.
\end{align}
Here the energy $E_n$ corresponds to the solution of (\ref{eq.8.11}) for vanishing order parameters, i.e. $\Psi^*=\Psi=0$, so we conclude $E_n=E_n^{(0)}$. With this (\ref{form.6.7}) coincides with the mean-field phase boundary (\ref{QWERT}). For the first two Mott lobes, we just set $n=1$ and $n=2$, which is depicted in FIG. \ref{fig.(1.99)}.

\subsubsection{Self-consistency equations}

Generally, the Brillouin-Wigner perturbation theory yields a polynomial representation of the ground-state energy $E_{n}$ and the condensate density $\Psi^{*} \Psi$ in orders of $\lambda$:
\begin{align}\label{eq.(1.6)}
0= A_0(E_n) + A_1(E_n,\Psi^* \Psi) \lambda + A_2(E_n,\Psi^* \Psi) \lambda^2 + \cdots .
\end{align}
By applying $\partial E_n/\left(\Psi \partial \Psi^{*}\right)=0$ to (\ref{eq.(1.6)}) we have
\begin{align}\label{eq.(1.7)}
0= &B_1(E_n,\Psi^* \Psi) + B_2(E_n,\Psi^* \Psi) \lambda+ B_3(E_n,\Psi^* \Psi) \lambda^2 + \cdots\, ,
\end{align}
with the coefficients
\begin{align}
B_{i}(E_n,\Psi^* \Psi) = \frac{1}{\Psi}\frac{\partial A_{i}(E_n,\Psi^* \Psi)}{\partial \Psi^{*}} \, .
\end{align}

Note that due to the derivative with respect to $\Psi^*$, the third-order coefficient $B_3$ appears in the second order of $\lambda$. The two equations (\ref{eq.(1.6)}) and (\ref{eq.(1.7)}) define both variables, i.e. the perturbed ground-state energy $E_{n}$ and the condensate density $\Psi^{*} \Psi$. Generically we have to solve them numerically in an iterative way. In order to get the energy and the condensate density within the one-state approach we calculate $\partial E_n/\left(\Psi \partial \Psi^{*}\right)=0$ from (\ref{eq.8.11}):
\begin{align}\label{eq.11.13}
0=&1 +\lambda J z \left(  \frac{n}{E_{n} - E_{n-1}^{(0)}} + \frac{n+1}{E_{n} - E_{n+1}^{(0)}}\right)+2\lambda ^2 J^2 z^2 \Psi^{*}\Psi \left[  \frac{n}{\left(E_{n} - E_{n-1}^{(0)}\right)^2} + \frac{n+1}{\left(E_{n} - E_{n+1}^{(0)}\right)^2}\right]\, ,
\end{align}
which corresponds to (\ref{eq.(1.7)}). Furthermore, by evaluating (\ref{eq.8.11}) up to second order in $\lambda$, we get
\begin{align}\label{eq.11.14}
0=&E_{n}^{(0)}-E_{n} + \lambda J z \Psi^{*}\Psi + \lambda ^2 J^2 z^2 \Psi^{*}\Psi \left(  \frac{n}{E_{n} - E_{n-1}^{(0)}} + \frac{n+1}{E_{n} - E_{n+1}^{(0)}}\right)
\,,
\end{align}
which corresponds to (\ref{eq.(1.6)}). Eliminating the denominators in (\ref{eq.11.13}) and (\ref{eq.11.14}) yields
\begin{align}\label{eq.11.15}
\nonumber
0=&\left(E_{n} - E_{n-1}^{(0)}\right)^2 \left(E_{n} - E_{n+1}^{(0)}\right)^2 +\lambda J z \left[  n\left(E_{n} - E_{n-1}^{(0)}\right)  \right. \left(E_{n} - E_{n+1}^{(0)}\right)^2 + \left. \left(n+1\right)\left(E_{n} - E_{n-1}^{(0)}\right)^2 \left(E_{n} - E_{n+1}^{(0)}\right)\right]
\\
& +2\lambda ^2 J^2 z^2 \Psi^{*}\Psi \left[  n\left(E_{n} - E_{n+1}^{(0)}\right)^2 \right. \left. +\left(n+1\right)\left(E_{n} - E_{n-1}^{(0)}\right)^2\right]
\end{align}
and
\begin{align}\label{eq.11.16}
0=&\left(E_{n} - E_{n-1}^{(0)}\right) \left(E_{n} - E_{n+1}^{(0)}\right) \left(E_{n}^{(0)}-E_{n} + \lambda J z \Psi^{*}\Psi \right) + \lambda ^2 J^2 z^2 \Psi^{*}\Psi \left[  n \left(E_{n} - E_{n+1}^{(0)}\right) \right. \left. +\left(n+1\right)\left(E_{n} - E_{n-1}^{(0)}\right) \right]
\,.
\end{align}
Both equations (\ref{eq.11.15}) and (\ref{eq.11.16}) are now used to calculate the ground-state energy $E_n$ and the condensate density $\Psi^* \Psi$. They are numerically solved by iteration.

\subsubsection{Energy and condensate density}
The energy is shown in Tab. \ref{tab.OnestateTab1}. At the degeneracy $\mu=U$, the unperturbed energy is given by $E_{n}^{(0)}=-U$. Therefore, the corrections of the energy in power series of $\lambda$ are obtained by subtracting the unperturbed energy from the perturbed energy. From zeroth to second order, the corrections amount to $+1.08\%$. From second to fourth order, the corrections are $-0.05\%$. Furthermore, from fourth to sixth order, the corrections are of the order $-0.18\%$. Note that for higher values of $Jz/U$ the convergence turns out to be slower.

\begin{table}[t]
\centering
    \begin{tabular}{cc|c|c|c|c}
\cline{3-5}
& & \multicolumn{3}{ c| }{Powers in $\lambda$} \\ \cline{3-5}
& & $\lambda^2$ & $\lambda^4$ & $\lambda^6$  \\ \cline{2-5}&
\multicolumn{1}{ |c| }{$\frac{E_{1}}{U}$} & -1.0108081 & -1.0102528 &  -1.0090297      \\ \cline{2-5}   
\end{tabular}
    \caption{Values for ground-state energy $E_{n}$ from the one-state approach at the degeneracy, i.e., $\mu=Un+\varepsilon$, $\varepsilon=0$, $\lambda =1$, $n=1$, and $Jz/U=0.02$. Columns give values for formulas evaluated up to second, fourth, and sixth order in $\lambda$.}\label{tab.OnestateTab1}   
\end{table}

\begin{figure}[t]
	\centering
\includegraphics[width=.6\textwidth]{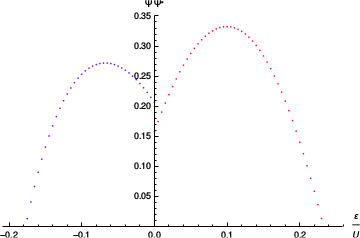}
\caption{(color online) Condensate density from one-state approach for $n=1$ (negative $\varepsilon/U$, purple) and $n=2$ (positive $\varepsilon/U$, red).} \label{fig.(11.2)}
\end{figure}

The condensate density $\Psi^* \Psi$ follows also from numerically solving both equations (\ref{eq.11.15}) and (\ref{eq.11.16}) iteratively. The result is plotted in FIG. \ref{fig.(11.2)} for $\mu = Un + \varepsilon$, $\lambda=1$ and $Jz/U=0.08$. We observe that the order parameter obtained from the Brillouin-Wigner perturbation theory for the one-state approach according to FIG. \ref{fig.(11.2)} is better than the one obtained from Rayleigh-Schr\"odinger perturbation theory, where the order parameter vanishes at the degeneracy as seen in FIG. \ref{fig.OPRS}. Nevertheless, the order parameter plotted in FIG. \ref{fig.(11.2)} still is discontinuous at $\varepsilon/U=0$ and thus does not yet represent a physically acceptable result. The resulting value for the order parameter is shown in Table \ref{tab.OnestateTab2} at the degeneracy $\mu=Un$ for $n=1$. Note that for higher values of $Jz/U$ the convergence is slower.
\\
\begin{table}[t]
\centering
    \begin{tabular}{cc|c|c|c|c}
\cline{3-5}
& & \multicolumn{3}{ c| }{Powers in $\lambda$} \\ \cline{3-5}
& & $\lambda^2$ & $\lambda^4$ & $\lambda^6$  \\ \cline{2-5}
&
\multicolumn{1}{ |c| }{$\Psi^{*}\Psi$} & 0.19862639 &  0.24896610 & 0.25601384 &     \\ \cline{2-5}
\end{tabular}
    \caption{Values for condensate density $\Psi^{*}\Psi$ from the one-state approach at the degeneracy, i.e., $\mu=Un+\varepsilon$, $\varepsilon=0$, $\lambda =1$, $n=1$, and $Jz/U=0.02$. Columns give values for formulas evaluated up to second, fourth, and sixth order in $\lambda$.}\label{tab.OnestateTab2}   
\end{table}

\subsubsection{Superfluid density}

Generally, the superfluid density is calculated by introducing a Galilei boost  \cite{Bradlyn}, which can be defined via
\begin{align}
E( \vec{\phi}) = E( \vec{0} ) + \rho_{\rm{SF}}N_{\rm{S}}\frac{1}{2}m \vec{v}^2 \,,
\end{align}
with $E( \vec{\phi})$ being the energy with a boost, $E( \vec{0} )$ the energy without a boost, $\rho_{\rm{SF}}$ is the superfluid density, $N_{\rm{S}}$ is the total number of sites, $m$ the mass, and $\vec{v}^2$ the velocity. Thus, we add a kinetic term to our energy, with the velocity
\begin{align}
\vec{v} = \frac{\hbar}{m}\frac{\vec{\phi}}{L}\,,
\end{align}
with $L$ denoting the spatial extend of the system in the direction of $\vec{v}$, where we have just introduced the velocity parameter $\vec{\phi}$. This adds an exponential term to the operators
\begin{align}
\hat{a}_j \to \hat{a}_j{\rm{e}}^{i \frac{\vec{x}_j}{L}\cdot \vec{\phi}} \,,
\end{align}
which finally gives rise to the substitution of the coordination number in the mean-field theory
\begin{align}
z \to  z - \left( \frac{a}{L} \right)^2 \vec{\phi}^2\,,
\end{align}
with $a$ the lattice spacing. Out of this, the superfluid density is determined as 
\begin{align}
\rho_{\rm{SF}}= \lim_{\vec{\phi} \to \vec{0}} \frac{2 L^2 \left[ E( \vec{\phi} ) - E( \vec{0} ) \right]}{ N_{\rm{S}}Ja^2\vec{\phi}^2} \,.
\end{align}
However, it is shown in Appendix \ref{appC}, that within the mean-field approximation the superfluid density always coincides with the condensate density. Thus, we conclude that the approximations within the mean-field approach are too strong to result in any difference between the condensate density and the superfluid density. In order to improve this, one must not apply the mean-field theory, but use some other method to deal with the system, like the field-theoretic method, where a Legendre transform of the grand-canonical free energy \cite{Ednilson,articleEdnilson,articleHolthaus} is used.

\subsection{Two-states approach}\label{Two-States Approach}

Now we consider the subspace of the Hilbert space which is spanned by $\ket{\Psi_{n}^{(0)}}$ and $\ket{\Psi_{n+1}^{(0)}}$. This choice is motivated due to the degeneracy present between two consecutive Mott lobes in the zero-temperature phase diagram of the Bose-Hubbard model. Any state vector is projected into that subspace by the projector
\begin{align}
\hat{P}=\ket{\Psi_{n}^{(0)}} \bra{\Psi_{n}^{(0)}} + \ket{\Psi_{n+1}^{(0)}} \bra{\Psi_{n+1}^{(0)}}\,,
\end{align}
and we will perform our calculations by evaluating (\ref{eq.2.37}) from the two-states approach.

\subsubsection{Quantum phase boundary}

\begin{figure}[t]
	\centering
\includegraphics[width=.6\textwidth]{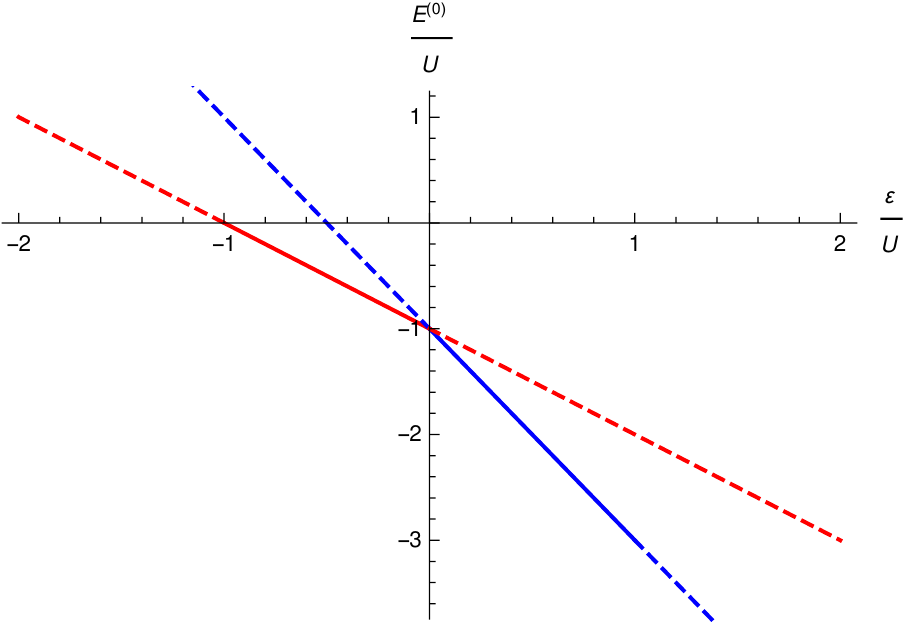}
\caption{(color online) The line with the smaller slope (red) corresponds to $E_{1}$ in (\ref{eq.(1.26)}), and the line with the bigger slope (blue) corresponds to $E_{2}$ in (\ref{eq.(1.27)}). The solid parts represent the lowest energies.} \label{fig.(1.14)}
\end{figure}

The mean-field quantum phase boundary was already shown in FIG. \ref{fig.vieleLobes} and FIG. \ref{fig.(1.99)}. In order to calculate the mean-field quantum phase boundary via the two-states approach, we start with the determinant of the matrix (\ref{eq.2.38}),
	\begin{align}\label{eq.(1.20)}
	\notag
	\rm{Det}\left(\Gamma\right)=&
	\left[\lambda^4 \frac{J^4 z^4 \Psi^{*2} \Psi^2 n \left( n-1 \right)}{ \left( E_{n} - E^{(0)}_{n-1}-\lambda Jz\Psi^*\Psi \right)^2 \left(  E_{n} - E^{(0)}_{n-2}-\lambda Jz\Psi^*\Psi \right) }+ E^{(0)}_{n}+\lambda Jz\Psi^*\Psi-E_{n}   \right.
	\\
	\notag
	& \left. +\lambda^2 \frac{J^2 z^2 \Psi^{*} \Psi n}{E_{n} - E^{(0)}_{n-1}-\lambda Jz\Psi^*\Psi} \right]  \left[ E^{(0)}_{n+1}+\lambda Jz\Psi^*\Psi-E_{n} +\lambda^2 \frac{J^2 z^2 \Psi^{*} \Psi \left( n+2 \right)}{E_{n} - E^{(0)}_{n+2}-\lambda Jz\Psi^*\Psi}\right.
	\\ 
	& \left. +\lambda^4 \frac{J^4 z^4 \Psi^{*2} \Psi^2 \left( n+2 \right) \left( n+3 \right)}{ \left(E_{n} - E^{(0)}_{n+2}-\lambda Jz\Psi^*\Psi \right)^2 \left(  E_{n} - E^{(0)}_{n+3}-\lambda Jz\Psi^*\Psi \right) }  \right] - \lambda^2 J^2 z^2\Psi^* \Psi \left(n+1\right) +... \,.
	\end{align}
To calculate the phase boundary we perform 
	\begin{align}\label{eq.(1.22)}
	\frac{1}{\Psi}\frac{\partial \rm{Det}\left(\Gamma\right)}{\partial\Psi^*}\Bigg|_{\Psi^* \Psi=0}=& \lambda Jz\left[\left( E_{n}^{(0)} -E_n\right)+\left(E_{n+1}^{(0)} -E_n \right)-\lambda Jz\left(n+1\right)\right]\nonumber\\
	&+\lambda^2 J^2 z^2\left[\frac{\left(n+2\right)\left(E_{n}^{(0)} -E_n\right)}{E_n - E_{n+2}^{(0)}}+\frac{n\left(E_{n+1}^{(0)} -E_n\right)}{E_n - E_{n-1}^{(0)}}\right] = 0\,,
	\end{align}
resulting in
	\begin{align}\label{1form.6.7}
	\frac{Jz}{U}=\frac{- \left( 2 E_{n} - E_{n}^{(0)}- E_{n+1}^{(0)} \right) \left( E_{n} - E_{n+2}^{(0)} \right) \left( E_{n} - E_{n-1}^{(0)} \right)}
	{ \lambda n U \left( E_{n} - E_{n+1}^{(0)} \right) \left( E_{n} - E_{n+2}^{(0)} \right) + \lambda U\left[ \left(n+1\right) \left( E_{n} - E_{n+2}^{(0)} \right)  + \left(n+2\right) \left( E_{n} - E_{n}^{(0)} \right)\right]}\,,
	\end{align}
which is the mean-field phase boundary. All higher order corrections drop out of the formula if we set $\Psi^* \Psi =0$. Thus, the phase boundary does not change even if higher orders in $\lambda$ are taken into account. To determine $E_n$ in (\ref{1form.6.7}), we take (\ref{eq.(1.20)}) and set $\Psi^* \Psi =0$, which results effectively in calculating the matrix up to zeroth order. We set it equal to zero,
\begin{align}
{\rm{Det}}\left(\Gamma\right)=\left( E_{n}^{(0)} -E_n \right)\left( E_{n+1}^{(0)} -E_n \right)=0\,,
\end{align}
and get two possibilities: $E_n=E_{n}^{(0)}$ or $E_n=E_{n+1}^{(0)}$. Thus, the mean-field phase boundary (\ref{1form.6.7}) with $\lambda=1$ agrees with the previous result (\ref{QWERT}). Using the explicit forms of the unperturbed energies (\ref{EN0uncurly}) together with $\mu = Un +\varepsilon$ for $n=1$, we have
\begin{align}\label{eq.(1.26)}
E_{1}=-\left(1+\frac{\varepsilon}{U}\right)U
\end{align}
and
\begin{align}\label{eq.(1.27)}
E_{2}=-\left(1+2\frac{\varepsilon}{U}\right)U\,.
\end{align}
These two energies are depicted in FIG. \ref{fig.(1.14)} and yield the lowest energies, corresponding to the two Mott lobes. For $-1<\varepsilon/U<0$, $E_1$ is the minimal energy, while for $0<\varepsilon/U<1$ it is $E_2$.

To get the phase boundary, we insert (\ref{eq.(1.26)}) and (\ref{eq.(1.27)}) into (\ref{1form.6.7}). According to FIG. \ref{fig.(1.14)}, $E_{1}$ gives rise to the first lobe, and $E_{2}$ to the second. Therefore, we obtain the Mott lobes in FIG. \ref{fig.(1.99)}, which were already discussed via FIG. \ref{fig.vieleLobes}.

\subsubsection{Energy and particle density}

We calculate the expectation value of the perturbed ground-state energy $E_{n}$ similarly to the previous section from the two conditions 
\begin{subequations}
\begin{align}
   \rm{Det}(\Gamma) &= 0 \,,\label{subeq1}\\
   \frac{1}{\Psi}\frac{\partial}{\partial \Psi^*}\rm{Det}(\Gamma)&=0\,,\label{subeq2}
\end{align}
\end{subequations}
where $\Gamma$ is given by
\begin{align}\label{eq.11.18}
\Gamma= \begin{pmatrix} E^{(0)}_{n}+\lambda J z \Psi^* \Psi-E_{n} +\lambda^2 \frac{J^2 z^2 \Psi^{*} \Psi n}{E_{n} - E^{(0)}_{n-1}-\lambda J z \Psi^* \Psi}
& -\lambda J z \Psi^{*} \sqrt{n+1} \\
-\lambda J z \Psi \sqrt{n+1}
& E^{(0)}_{n+1}+\lambda J z \Psi^* \Psi-E_{n} +\lambda^2 \frac{J^2 z^2 \Psi^{*} \Psi \left( n+2 \right)}{E_{n} - E^{(0)}_{n+2}-\lambda J z \Psi^* \Psi}
\end{pmatrix}\,.
\end{align}
The perturbed ground-state energy $E_n$ is then determined by solving both equations (\ref{subeq1}) and (\ref{subeq2}) iteratively.

\begin{table}[t]
  \centering
  \begin{tabular}{ |l|l| }
  \hline
  $\frac{Jz}{U}=0.02$ & $\frac{E_n}{U}=-56.6 \left( \frac{\varepsilon}{U} \right)^6-3.1 \left( \frac{\varepsilon}{U} \right)^5+7.5 \left( \frac{\varepsilon}{U} \right)^4+3.1 \left( \frac{\varepsilon}{U} \right)^3-6.0 \left( \frac{\varepsilon}{U} \right)^2-1.5 \frac{\varepsilon}{U} -1.0$ \\ \hline
  $\frac{Jz}{U}=0.08$ & $\frac{E_n}{U}=-1.4 \left( \frac{\varepsilon}{U} \right)^6-0.8 \left( \frac{\varepsilon}{U} \right)^5+1.7 \left( \frac{\varepsilon}{U} \right)^4+0.7 \left( \frac{\varepsilon}{U} \right)^3-1.4 \left( \frac{\varepsilon}{U} \right)^2-1.5  \frac{\varepsilon}{U}-1.0$ \\ \hline
   $\frac{Jz}{U}=0.101$ & $\frac{E_n}{U}=-0.6 \left( \frac{\varepsilon}{U} \right)^6-0.8 \left( \frac{\varepsilon}{U} \right)^5+1.2 \left( \frac{\varepsilon}{U} \right)^4+0.5 \left( \frac{\varepsilon}{U} \right)^3-1.1 \left( \frac{\varepsilon}{U} \right)^2-1.5 \frac{\varepsilon}{U} -1.1$ \\
  \hline
\end{tabular}
    \caption{Fit functions for $E_n/U$ for the three different values of $Jz/U$, according to FIG. \ref{Energies} (b).} \label{tableIII}
\end{table}

\begin{figure}[t]
   \centering
   \begin{subfigure}{0.47\columnwidth}
      \includegraphics[width=\columnwidth]{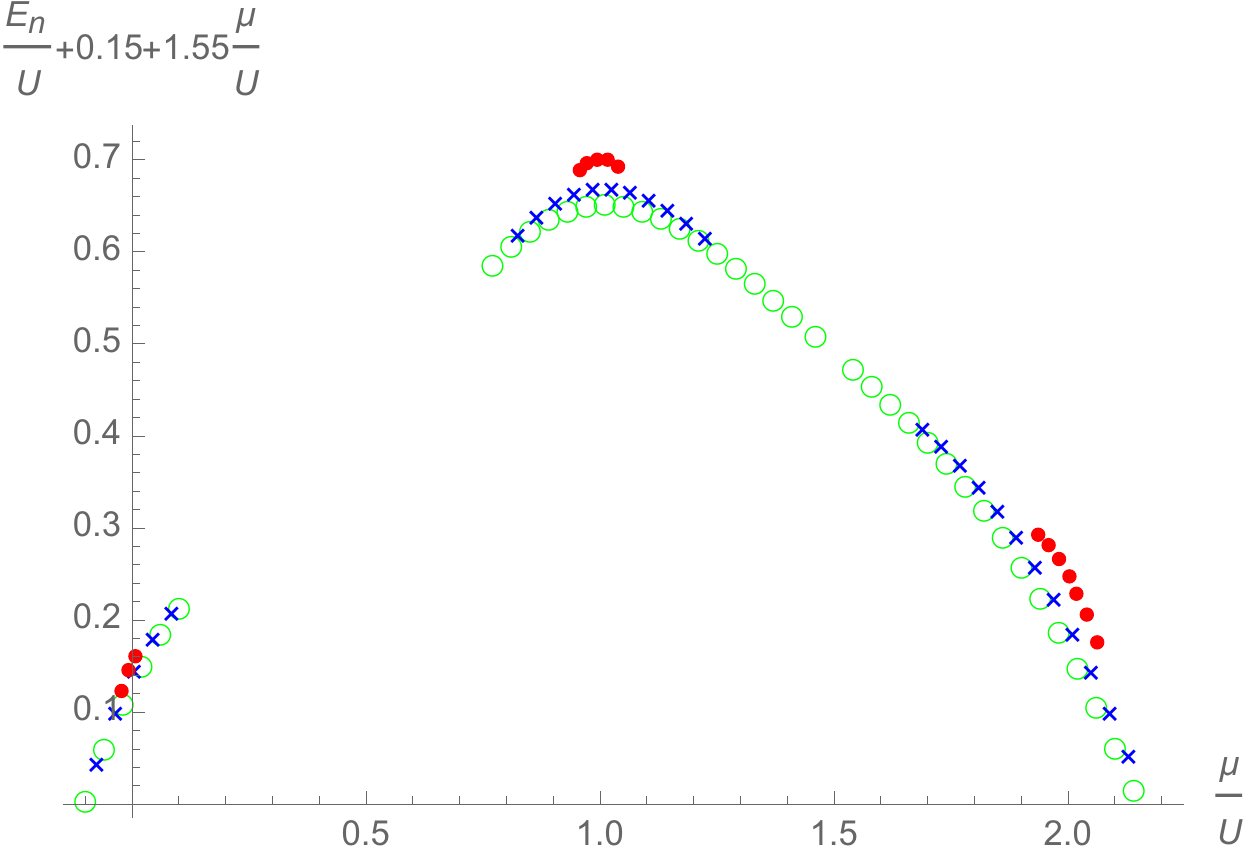}
        \caption{The energy in the superfluid phase is shown for the first two Mott lobes. The central part is shown in FIG. \ref{Energies} (b). For better visualization, the linear equation $0.15+1.55 \mu/U$, which scales the outmost points of the green plot to zero, is added to the energy.}
        \label{fig:1.1a}
   \end{subfigure}
   \qquad
  \begin{subfigure}{0.47\columnwidth}
       \includegraphics[width=\columnwidth]{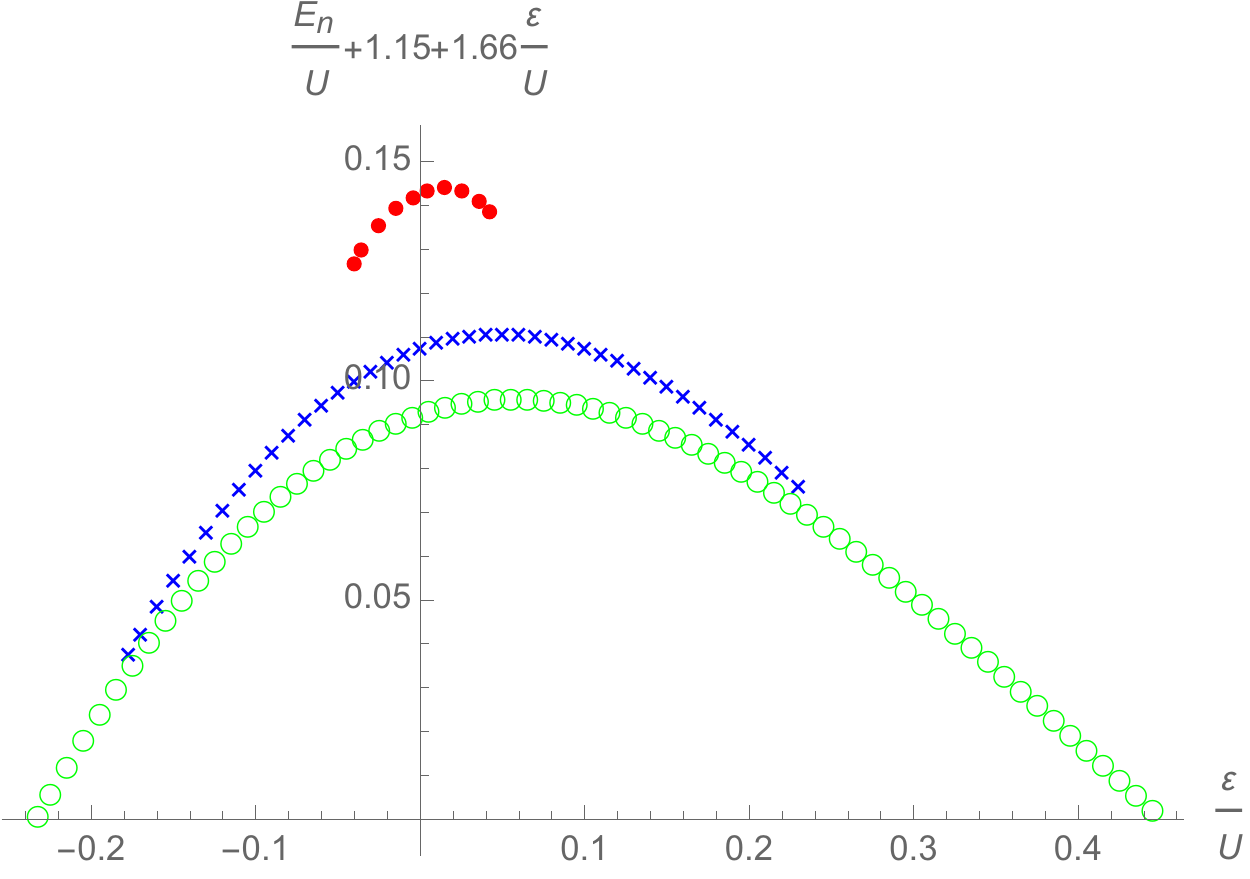}
       \caption{The energy in between the lobes for $n=1$ and $n=2$, centered around the degeneracy by introducing $\mu=Un+\varepsilon$. For better visualization, the linear equation $1.15+1.66 \varepsilon/U$, which scales the outmost points of the green plot to zero, is added to the energy.}
        \label{fig:1.1b}
    \end{subfigure}
    \caption{(color online) The energy for all the superfluid parts of the lines  with the respective colors in FIG. \ref{fig.vieleLobes}. Perturbed ground-state energies $E_n/U$ up to $\lambda^4$ between the Mott lobes in the superfluid region for different values of $Jz/U$: $Jz/U=0.02$ (dots, red), $Jz/U=0.08$ (crosses, blue) and $Jz/U=5-2\sqrt{6} \approx 0.101$ (rings, green). At $Jz/U=5-2\sqrt{6}$ the second lobe hits exactly its tip.}\label{Energies}
\end{figure}

\begin{table}[t]
	\centering
	\begin{subtable}{\linewidth}
		\begin{tabular}{ |l|l| }
			\hline
			$0 \leq n \leq 1$ & $\frac{E_n}{U}=-0.005- 0.5 \frac{\mu}{U} - 12 \left(\frac{\mu}{U}\right)^2 + 12 \left(\frac{\mu}{U}\right)^3 + 3 \left(\frac{\mu}{U}\right)^4 + 3 \left(\frac{\mu}{U}\right)^5 - 402 \left(\frac{\mu}{U}\right)^6$ \\ \hline
			$1 \leq n \leq 2$ & $\frac{E_n}{U}=-54 + 314 \frac{\mu}{U} -  788 \left(\frac{\mu}{U}\right)^2 + 1074 \left(\frac{\mu}{U}\right)^3 -  826 \left(\frac{\mu}{U}\right)^4 + 336 \left(\frac{\mu}{U}\right)^5 -  57 \left(\frac{\mu}{U}\right)^6$ \\ \hline
			$2 \leq n \leq 3$ & $\frac{E_n}{U}=-1297 + 3963 \frac{\mu}{U} -  5053 \left(\frac{\mu}{U}\right)^2 + 3429 \left(\frac{\mu}{U}\right)^3 - 1305 \left(\frac{\mu}{U}\right)^4 + 264 \left(\frac{\mu}{U}\right)^5 -  22 \left(\frac{\mu}{U}\right)^6$ \\
			\hline
		\end{tabular}
		\caption{$Jz/U=0.02$ (red line).}\label{Fitsfor002} 
	\end{subtable}\par
	\begin{subtable}{\linewidth}
		\begin{tabular}{ |l|l| }
			\hline
			$0 \leq n \leq 1$ & $\frac{E_n}{U}=-0.02 - 0.5 \frac{\mu}{U} - 3\left(\frac{\mu}{U}\right)^2 + 3 \left(\frac{\mu}{U}\right)^3 + 
			1 \left(\frac{\mu}{U}\right)^4 + 1 \left(\frac{\mu}{U}\right)^5 -  18 \left(\frac{\mu}{U}\right)^6$ \\ \hline
			$1 \leq n \leq 2$ & $\frac{E_n}{U}=-0.5 + 1 \frac{\mu}{U} - 
			6 \left(\frac{\mu}{U}\right)^2 + 14 \left(\frac{\mu}{U}\right)^3 -  15\left(\frac{\mu}{U}\right)^4 +8 \left(\frac{\mu}{U}\right)^5 - 
			1 \left(\frac{\mu}{U}\right)^6$ \\ \hline
			$2 \leq n \leq 3$ & $\frac{E_n}{U}=0.7 + 13 \frac{\mu}{U} -  38 \left(\frac{\mu}{U}\right)^2 + 37 \left(\frac{\mu}{U}\right)^3 - 
			18 \left(\frac{\mu}{U}\right)^4 + 4 \left(\frac{\mu}{U}\right)^5 -  0.4 \left(\frac{\mu}{U}\right)^6$ \\
			\hline
		\end{tabular}
		\caption{$Jz/U=0.08$ (blue line).} \label{Fitsfor008}
	\end{subtable}
	\begin{subtable}{\linewidth}
		  \begin{tabular}{ |l|l| }
		  	\hline
		  	$0 \leq n \leq 1$ & $\frac{E_n}{U}=-0.03 - 0.5 \frac{\mu}{U} - 2 \left(\frac{\mu}{U}\right)^2 + 2 \left(\frac{\mu}{U}\right)^3 + 
		  	0.8 \left(\frac{\mu}{U}\right)^4 + 0.7 \left(\frac{\mu}{U}\right)^5 -  10 \left(\frac{\mu}{U}\right)^6$ \\ \hline
		  	$1 \leq n \leq 2$ & $\frac{E_n}{U}=0.2 - 3 \frac{\mu}{U} + 3 \left(\frac{\mu}{U}\right)^2 - 0.2 \left(\frac{\mu}{U}\right)^3 - 
		  	4 \left(\frac{\mu}{U}\right)^4 + 3 \left(\frac{\mu}{U}\right)^5 -  0.6 \left(\frac{\mu}{U}\right)^6$ \\ \hline
		  	$2 \leq n \leq 3$ & $\frac{E_n}{U}=-24 + 86 \frac{\mu}{U} - 130 \left(\frac{\mu}{U}\right)^2 + 99 \left(\frac{\mu}{U}\right)^3 - 
		  	41 \left(\frac{\mu}{U}\right)^4 + 9 \left(\frac{\mu}{U}\right)^5 -  0.8 \left(\frac{\mu}{U}\right)^6$ \\
		  	\hline
		  \end{tabular}
		  \caption{$Jz/U=0.101$ (green line).} \label{Fitsfor0101}
	\end{subtable}
	\caption{Fit functions for $E_n/U$  in FIG. \ref{fig:002,008,0101PD} corresponding to three different hopping values in between the different Mott lobes.}
\end{table}

\begin{figure}[t]
	\centering
	\begin{subfigure}{0.47\textwidth}
		\includegraphics[width=\textwidth]{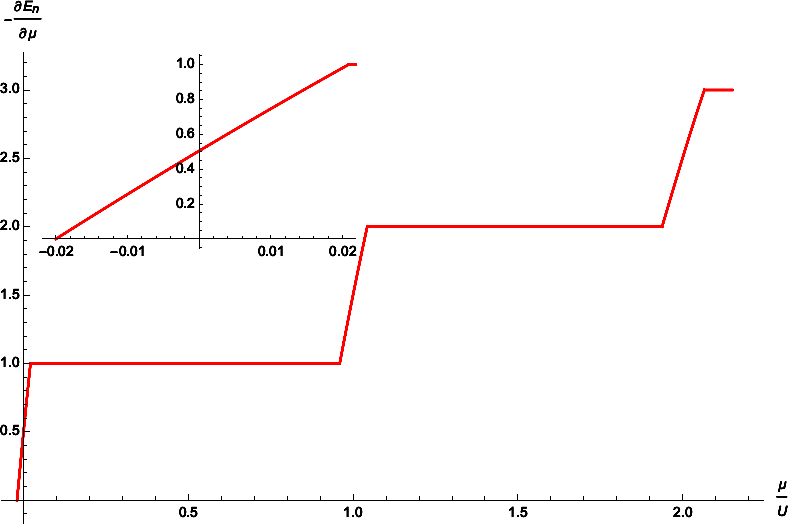}
		\caption{$Jz/U=0.02$.}
		\label{fig:002a}
	\end{subfigure}
	~ 
	\begin{subfigure}{0.47\textwidth}
		\includegraphics[width=\textwidth]{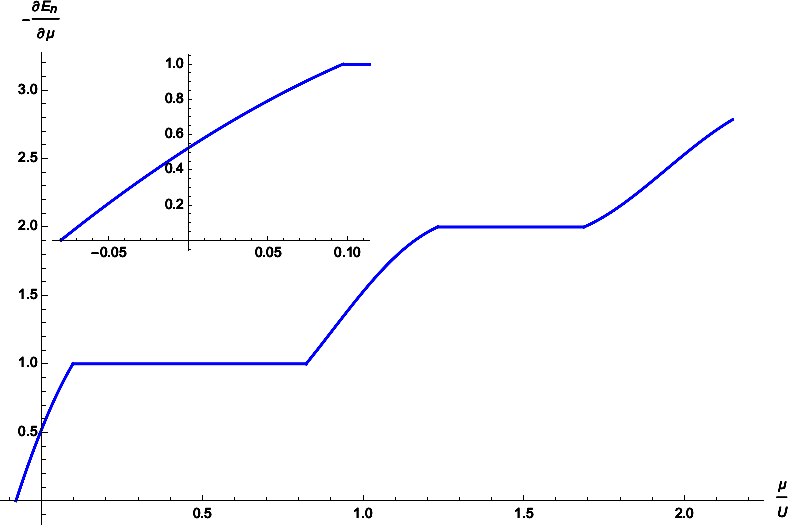}
		\caption{$Jz/U=0.08$.}
		\label{fig:008b}
	\end{subfigure}
	~ 
	\begin{subfigure}{0.47\textwidth}
		\includegraphics[width=\textwidth]{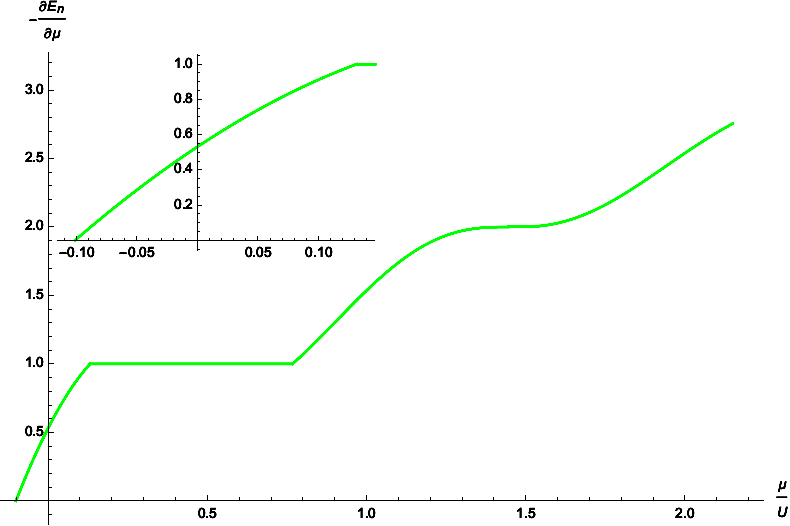}
		\caption{$Jz/U=0.101$.}
		\label{fig:0101}
	\end{subfigure}
	\caption{(color online) Particle density $-\partial E_n/\partial \mu$ over chemical potential $\mu/U$ according to corresponding lines in FIG. \ref{fig.vieleLobes}. Horizontal lines are within the Mott lobes, while curves are in the superfluid. For higher values of $Jz/U$, the curves become rounder.}\label{fig:002,008,0101PD}
\end{figure}

The plots in FIG. \ref{Energies} correspond to $\lambda=1$ considering the fit functions from Table \ref{tableIII}. The distance between two points is $\varepsilon/U = 0.005$. The ground-state energy $E_n$ is depicted as a function of the chemical potential for the superfluid regions, i.e. between Mott lobes, which explains the missing points in some regions in FIG. \ref{Energies}.

In order to get the particle density, shown in FIG. \ref{fig:002,008,0101PD}, we have to combine FIG. \ref{fig.vieleLobes} with FIG. \ref{Energies} (a). We do this exemplarily for the value of $Jz/U=0.02$, which is depicted by the first line from the bottom (red) in FIG. \ref{fig.vieleLobes}. Starting from the left, at zero particles $n=0$, we can read off that we are in the superfluid region. Thus, we take the numerical data for the energy in the superfluid region from FIG. \ref{fig:002,008,0101PD}, and fit them with a polynomial. This is done for the different superfluid regions and for different $Jz/U$ in Tab. \ref{Fitsfor002}--\ref{Fitsfor0101}. We calculate $-\partial E/\partial \mu$ to get the particle density in the superfluid region, which is plotted in FIG. \ref{fig:002,008,0101PD}. In the Mott lobes, whose boundaries can be read off from FIG. \ref{fig.vieleLobes}, we have a constant particle number, and thus a horizontal line, according to the particle number in the lobes in FIG. \ref{fig.vieleLobes}. In Tab. \ref{tab.TwostateTabEner}, the numerical value for the energy at the degeneracy $\mu=Un$ is shown. Note that for higher values of $Jz/U$, the convergence is slower.

\begin{table}[t]
\centering
    \begin{tabular}{cc|c|c|c|c}
\cline{3-5}
& & \multicolumn{3}{ c| }{Powers in $\lambda$} \\ \cline{3-5}
& & $\lambda^2$ & $\lambda^4$ & $\lambda^6$  \\ \cline{2-5}&
\multicolumn{1}{ |c| }{$\frac{E_{1}}{U}$} & -1.0100015 & -1.0104087  & -1.0104088      \\ \cline{2-5}   
\end{tabular}
    \caption{Values for ground-state energy $E_{n}$ from the two-states approach at the degeneracy, i.e., $\mu=Un+\varepsilon$, $\varepsilon=0$, $\lambda =1$, $n=1$, and $Jz/U=0.02$. Columns give values for formulas evaluated up to second, fourth, and sixth order in $\lambda$. }\label{tab.TwostateTabEner}   
\end{table}

\subsubsection{Condensate density}

The corresponding results for the condensate density $\Psi^* \Psi$ are plotted in FIG. \ref{fig:002,008,0101OP} and FIG. \ref{20OPS}, where we have set $\mu = Un + \varepsilon$, $\lambda=1$ and $n=1$. The distance between two points is $\varepsilon/U = 0.005$. The graphs corresponding to the condensate density have a maximum at $\varepsilon/U>0$ and they always go from the phase boundary of the Mott lobe with $n=1$ up to the phase boundary of the Mott lobe with $n=2$. Note that these different values for $n$ are already taken into account by the structure of the matrix (\ref{eq.11.18}), therefore we evaluate the whole matrix with the numerical value $n=1$, but get the physical result for the right half of the Mott lobe $n=1$ and for the left half of the Mott lobe we have to put $n=2$.

\begin{figure}[t]
	\centering
	\begin{subfigure}{.47\textwidth}
		\includegraphics[width=\textwidth]{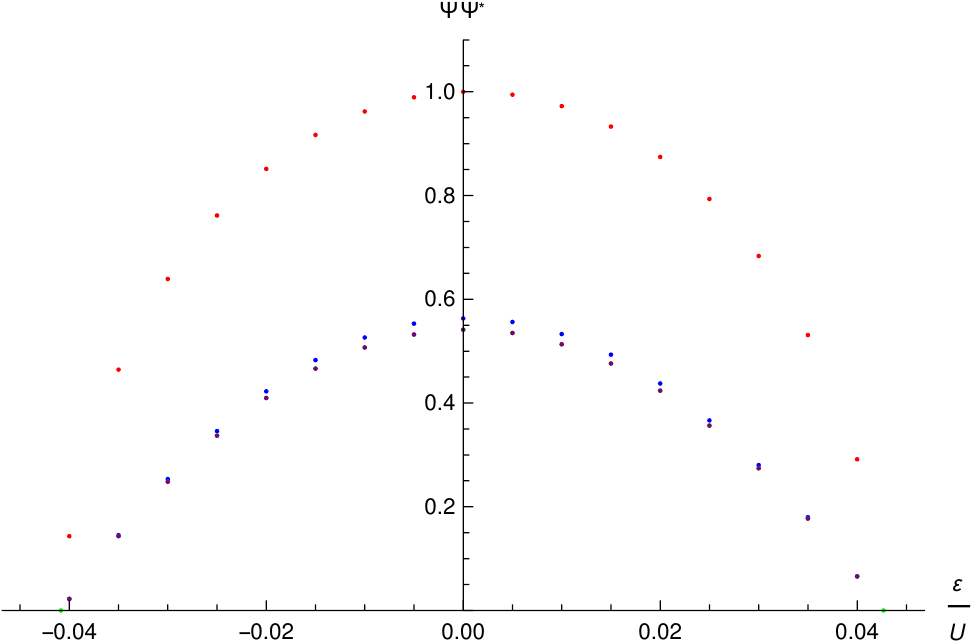}
		\caption{$Jz/U=0.02$.}
		\label{fig:002aa}
	\end{subfigure}
	~ 
	\begin{subfigure}{.47\textwidth}
		\includegraphics[width=\textwidth]{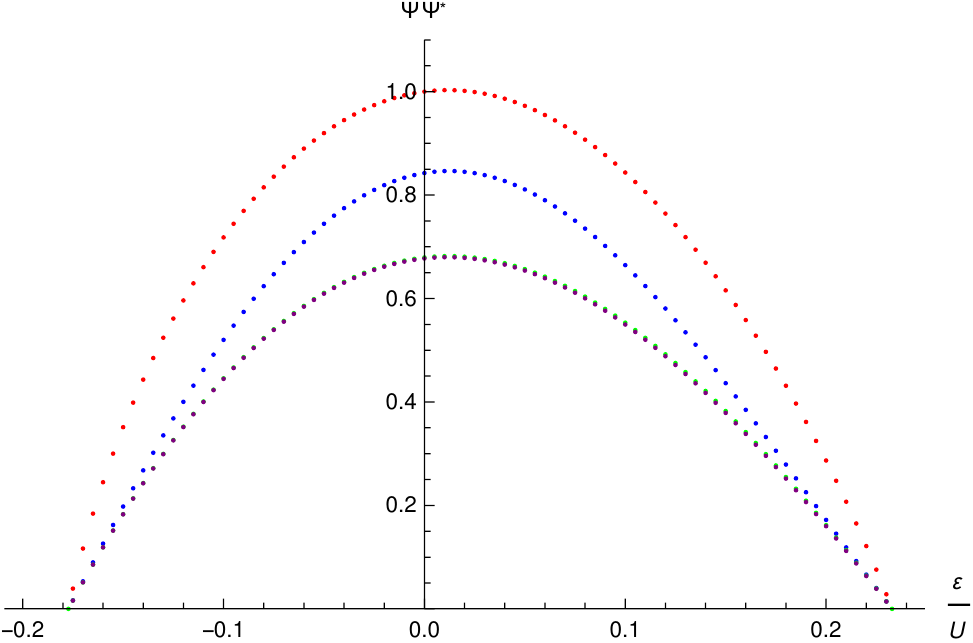}
		\caption{$Jz/U=0.08$.}
		\label{fig:008bb}
	\end{subfigure}
	~ 
	\begin{subfigure}{.47\textwidth}
		\includegraphics[width=\textwidth]{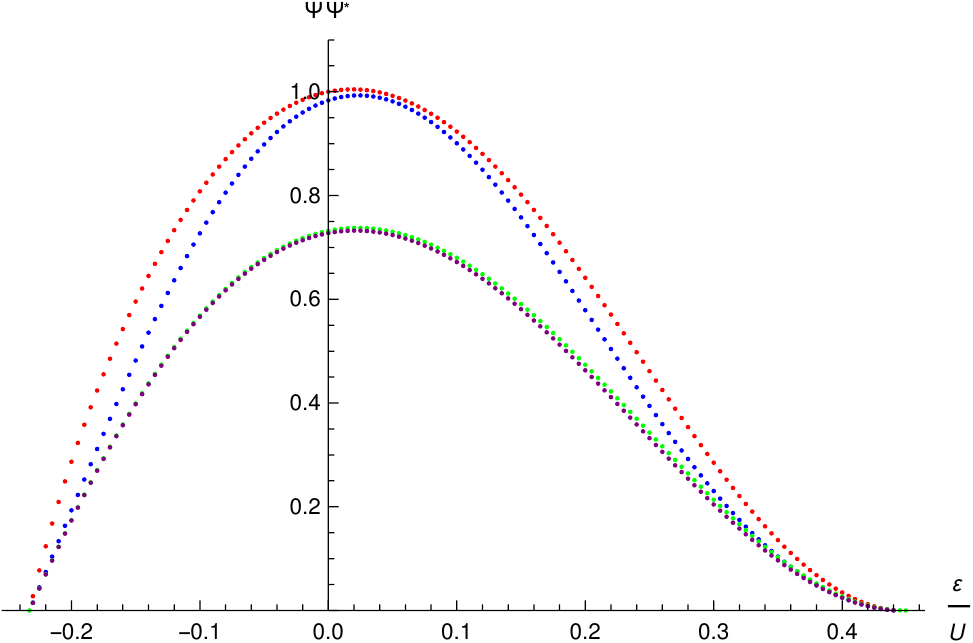}
		\caption{$Jz/U=0.101$.}
		\label{fig:0101c}
	\end{subfigure}
	\caption{(color online) Condensate density as a function of $\varepsilon/U=\mu/U-n$ for $\lambda=1$ and $n=1$. In each plot, the curves from the top to the bottom correspond to corrections up to the order $\lambda$ (red), $\lambda^2$ (blue), $\lambda^3$ (green), and $\lambda^4$ (purple). For small values of $Jz/U$ and thus close to the degeneracy, like in (a), the third (green) and fourth (purple) curves coincide.}\label{fig:002,008,0101OP}
\end{figure}

\begin{figure}[t]
	\centering
\includegraphics[width=.6\textwidth]{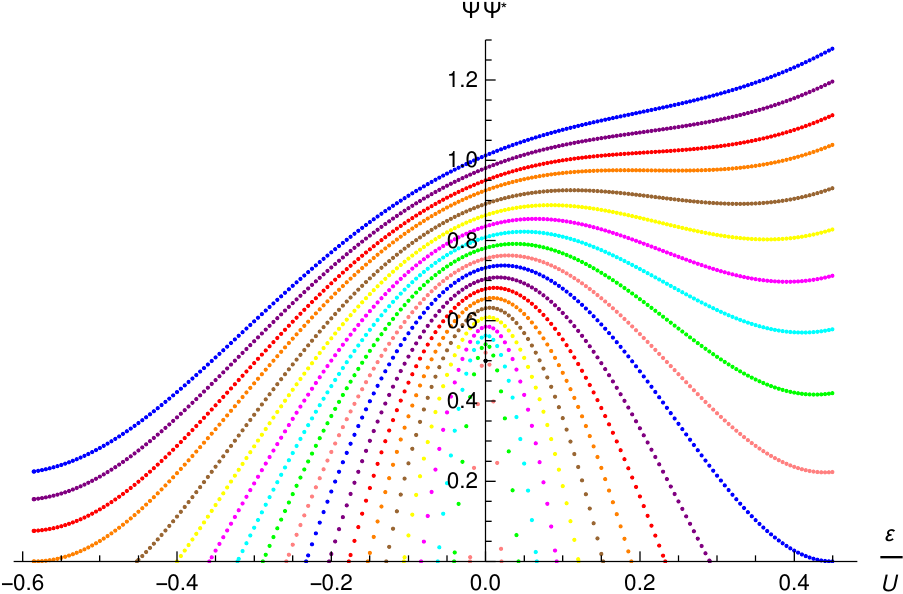}
\caption{(color online) Condensate density $\Psi^{*} \Psi$ as a function of $\varepsilon/U=\mu/U-n$ for $\lambda=1$ and $n=1$ up to $\lambda^4$ between the Mott lobes for different values of $Jz/U$, between $Jz/U=0$ and $Jz/U=0.20$ with a step size of 0.01 for $Jz/U$.}\label{20OPS}
\end{figure}

\begin{table}[t]
	\centering
	\begin{tabular}{cc|c|c|c|c}
		\cline{3-5}
		& & \multicolumn{3}{ c| }{Powers in $\lambda$} \\ \cline{3-5}
		& & $\lambda^2$ & $\lambda^4$ & $\lambda^6$  \\ \cline{2-5}&
		\multicolumn{1}{ |c| }{$\Psi^{*}\Psi$} & 0.56303521 &  0.54132128 & 0.54131277     \\ \cline{2-5}   
	\end{tabular}
	\caption{Values for the condensate density $\Psi^* \Psi$ from the two-states approach at the degeneracy, i.e. $\mu=Un+\varepsilon$, $\varepsilon=0$, $\lambda =1$, $n=1$, and $Jz/U=0.02$. Columns give values for formulas evaluated up to second, fourth, and sixth order in $\lambda$. For higher values of $Jz/U$ the convergence is slower. }\label{tab.1.1}  
\end{table}

FIG. \ref{fig:002,008,0101OP} shows different plots of the condensate density $\Psi^* \Psi$ over $\varepsilon/U$. There, it is depicted in a graphical way that the results converge for higher orders in $\lambda$. This is shown numerically in TAB. \ref{tab.1.1}. There we also see in comparison with Table \ref{tab.OnestateTab2} that the two-states approach converges faster than the one-state approach. Furthermore, the difference of the condensate density from the two-states approach in $\lambda^4$ to $\lambda^6$ is about $ 0.0016 \% $, which justifies to truncate the perturbative series already at fourth order in $\lambda$.

In Tab. \ref{tab.CDFits}, the fit functions for the condensate density is shown up to fourth order in $\lambda$, which corresponds to the fourth curve starting from the top (purple) in FIG. \ref{fig:002,008,0101OP}. Note that for $\varepsilon=0$, i.e. at the degeneracy, the condensate density does not always possess the same value, neither is it zero, as it was in FIG. \ref{fig.OPRS}.

FIG. \ref{20OPS} illustrates the condensate density $\Psi^* \Psi$ over $\varepsilon/U$ for 20 different values of $Jz/U$. For $Jz/U=0$, we get the black point at $\Psi^* \Psi =0.5$. For $Jz/U=0.01$ (pink) up to $Jz/U=0.09$ (purple) we get an approximately parabola shaped graph. For $Jz/U=5-2\sqrt{6} \approx 0.101$ (blue), we hit the second Mott lobe at its tip, and the graph touches the $\varepsilon/U$-axis in just one point for positive $\varepsilon/U$. For $Jz/U=0.11$ (pink) up to $Jz/U=0.16$, the part of the graph with positive $\varepsilon/U$ has still a minimum, while the negative parts intersect the $\varepsilon/U$-axis. For $Jz/U=3 - 2 \sqrt{2} \approx 0.172$ (orange), which is the tip of the first lobe, the part for negative $\varepsilon/U$ touches the $\varepsilon/U$-axis. For $Jz/U=0.18$ (red) up to $Jz/U=0.20$ (blue), which is just in the superfluid phase without touching any phase boundary, the whole graph is monotonically increasing. Note that this is a representation of the condensate density $\Psi^* \Psi$ which gives a non-zero, continuous result at the degeneracy, which was neither obtained by the Rayleigh-Schrödinger perturbation theory (see FIG. \ref{fig.OPRS}) \cite{thesisHoffmann} nor by the Brillouin-Wigner one-state approach (see FIG. \ref{fig.(11.2)}) \cite{articleMelo}. Therefore, for future calculations, the condensate density out of the Brillouin-Wigner two-states matrix approach should be used.

	\begin{table}[t]
		\centering
		\begin{tabular}{ |l|l| }
			\hline
			$\frac{Jz}{U}=0.02$ & $\Psi^* \Psi=0.54 + 0.29 \frac{\varepsilon}{U} - 311.46 \left( \frac{\varepsilon}{U} \right)^2 + 156.85 \left( \frac{\varepsilon}{U} \right)^3 + 387.87 \left( \frac{\varepsilon}{U} \right)^4 - 
			149.16 \left( \frac{\varepsilon}{U} \right)^5 - 2980.18 \left( \frac{\varepsilon}{U} \right)^6$ \\ \hline
			$\frac{Jz}{U}=0.08$ & $\Psi^* \Psi=0.68 + 0.44 \frac{\varepsilon}{U} - 18.28 \left( \frac{\varepsilon}{U} \right)^2 + 10.38 \left( \frac{\varepsilon}{U} \right)^3 + 29.03 \left( \frac{\varepsilon}{U} \right)^4 - 
			7.37 \left( \frac{\varepsilon}{U} \right)^5 - 6.17 \left( \frac{\varepsilon}{U} \right)^6$ \\ \hline
			$\frac{Jz}{U}=0.101$ & $\Psi^* \Psi=0.73 + 0.49 \frac{\varepsilon}{U} - 10.98 \left( \frac{\varepsilon}{U} \right)^2 + 6.47 \left( \frac{\varepsilon}{U} \right)^3 + 19.99 \left( \frac{\varepsilon}{U} \right)^4 - 
			2.47 \left( \frac{\varepsilon}{U} \right)^5 - 11.17 \left( \frac{\varepsilon}{U} \right)^6$ \\ 
			\hline
		\end{tabular}
		\caption{Fit functions for $\Psi^* \Psi$ for three different values of $Jz/U$ in fourth order in $\lambda$.}\label{tab.CDFits} 
	\end{table}

\subsection{Comparison between one-state approach, two-states approach, and numerics}

\begin{figure}[t]
	\centering
	\includegraphics[width=.6\textwidth]{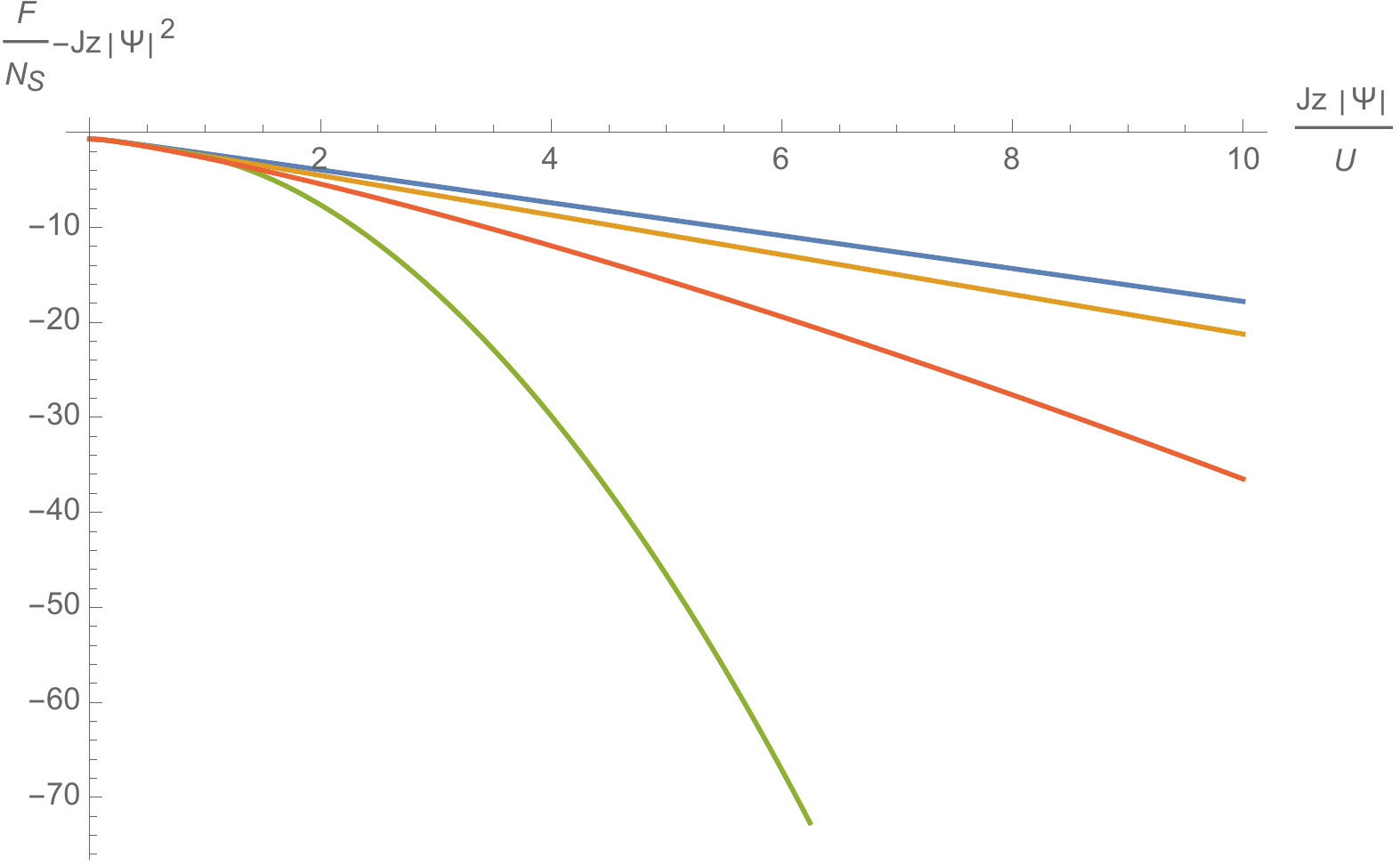}
	\caption{(color online) Ground-state energy $E_1$ out of one-state approach for $\mu=0.7U$. From the top to the bottom the respective curve represent the exact numerical value (blue) as well as the corrections $\lambda^6$ (yellow), $\lambda^4$ (red), and $\lambda^2$ (green). The labeling of the axis is motivated from (\ref{meanfieldfreeeergy}).}\label{NumVerg1}
\end{figure}

By comparing our analytic approach with purely numeric results, obtained by direct numerical diagonalization, we find a good convergence for small $Jz/U$. In FIG. \ref{NumVerg1}, the first curve from the top (blue) stems from the purely numeric calculation, while the other curves are from the one-state approach. The three curves are, starting from the bottom, up to $\lambda^2$ (green), $\lambda^4$ (red), and $\lambda^6$ (yellow). Thus, for small values of $Jz/U$, the one-state energy is quasi-exact. By comparing Tab. \ref{tab.OnestateTab1} with Tab. \ref{tab.TwostateTabEner}, we see that the energies from the one-state and the two-states approach coincide. Therefore, the two-states approach can be considered as well quasi-exact at least concerning the ground-state energy.

\section{\label{Trap}Trap}

In view of actual experiments, we consider now the impact of the harmonic confinement upon the equation of state. Although most traps in experiments have an ellipsoidal shape, we perform here calculations for the case of a spherical trap. In order to add a trap to our calculations, we have to perform the Thomas-Fermi, or local density approximation \cite{bookSmith,bookStringari}
\begin{align}\label{muTRAP}
\mu = \tilde{\mu}-\frac{1}{2}m\omega^2 |\vec{r}|^2\,.
\end{align}
Here, $m$ denotes the mass of the particles and $\omega$ stands for the trap frequency. Thus, the chemical potential is now consisting of a trap term and the original chemical potential $\tilde{\mu}$.

This procedure effectively gives rise to the same picture as in FIG. \ref{fig:002,008,0101PD}. We identify $\tilde{\mu}_{max}$ with the center of the trap, while the border of the trap is identified with the vanishing point of the condensate density. In between, we have Mott-insulating and superfluid regions, which give, in a three-dimensional trap, a wedding-cake structure with alternating Mott-insulating and superfluid shells.

In order to identify one of the graphs from FIG. \ref{fig:002,008,0101PD} with an actual experimental setting for a trap, we have to determine $\tilde{\mu}$. This is done by integrating over the plots from FIG. \ref{fig:002,008,0101PD}. Doing so results in a gauge curve for the equation of state for the total particle number, which allows to determine the corresponding value for $\tilde{\mu}$. 

\begin{figure}[t]
   \centering
      \includegraphics[width=.6\columnwidth]{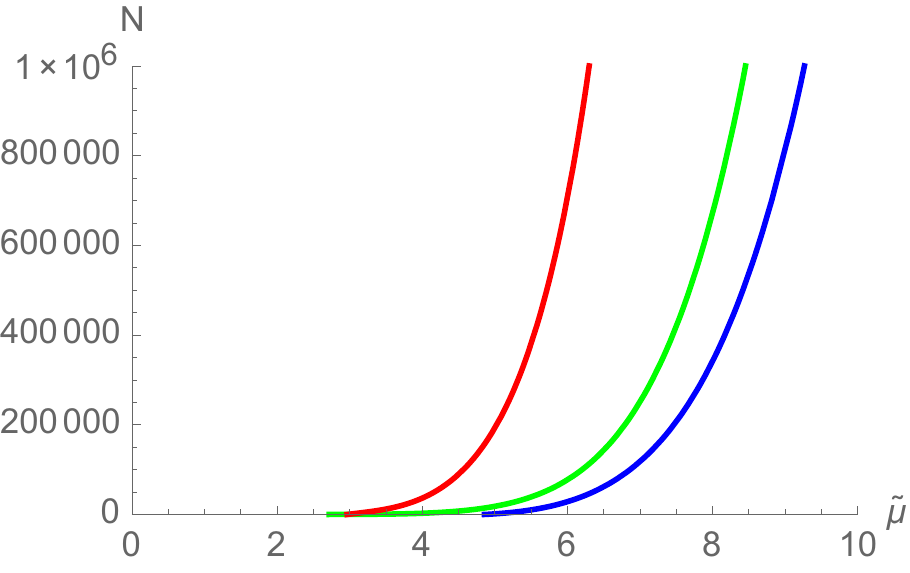}
        \caption{(color online) Equations of state $N=N(\tilde{\mu})$, with $m=87u$, $a=400 \rm{nm}$, and $\omega=48\pi \rm{Hz}$. From left to right: $Jz/U=0.02$ (red), $Jz/U=0.101$ (green), and $Jz/U=0.08$ (blue).}
\label{fig.Imu008}
\end{figure}

At first, we write down the integral and switch from Cartesian to spherical coordinates and perform the angular integrations
\begin{align}
I_{\mu_i, \mu_o}=-\frac{1}{a^3} \int\limits_{V}^{}\frac{\partial E_n}{\partial \mu} dV = -\frac{4\pi}{a^3} \int\limits_{R_i}^{R_o}r^2\frac{\partial E_n}{\partial \mu}dr\,,
\end{align}
where the radii $R_i$ and $R_o$ are the inner and the outer radius of the shell we want to compute, respectively. The further calculations are done for $Jz/U=0.08$ and $2 \leq n \leq 3$ (see FIG. \ref{fig:002,008,0101PD} (b), $1.69 \leq \mu/U \leq 2.15$), which is just the innermost superfluid shell. To this end we use the fit function for the energy in this region from Tab. \ref{Fitsfor008} and execute the differentiation:
\begin{align}\label{Integral1D}
I_{1.69, 2.15}=&-\frac{4\pi}{a^3} \int\limits_{R_2}^{R_3}r^2\left[13 - 38 \frac{\mu}{U} + 37 \left(\frac{\mu}{U}\right)^2  - 18 \left(\frac{\mu}{U}\right)^3 + 4 \left(\frac{\mu}{U}\right)^4 -  0.4 \left(\frac{\mu}{U}\right)^5 \right]dr\,,
\end{align}
with
\begin{subequations}
	\begin{align}
		R_3 &= \sqrt{\frac{2(\tilde{\mu}-2.15U)}{m\omega^2}} \, ,\\
		R_2 &= \sqrt{\frac{2(\tilde{\mu}-1.69U)}{m\omega^2}} \, .
	\end{align}
\end{subequations}
The last step is to insert (\ref{muTRAP}) into (\ref{Integral1D}) and perform the integration. The same procedure has to be repeated for all the other regions in FIG. \ref{fig:002,008,0101PD} (b), namely $I_{1.23, 1.69}$, $I_{0.82, 1.23}$, $I_{0.10, 0.82}$, and $I_{-0.08, 0.10}$, which represent the other superfluid and Mott insulating shells, respectively. These equations have to be added together in order to obtain the total particle number

\begin{align}
N=&I_{-0.08, 0.10}+I_{0.10, 0.82}+I_{0.82, 1.23}+I_{1.23, 1.69}+I_{1.69, 2.15}\,.
\end{align}
 The plot of the resulting equation of state $N=N(\tilde{\mu})$ is shown in FIG. \ref{fig.Imu008}. For small values of $\tilde{\mu}$, the particle number decreases to zero. From this plot, we conclude that for a given $\tilde{\mu}$ the minimal particle number is not at $Jz/U=0$, where all particles are in the Mott-insulator phase, neither is at $Jz/U > 0.172$, where all particles are in the superfluid phase. Instead, the minimal particle number is achieved for a specific distribution of Mott-insulator and superfluid, represented by a corresponding value of $Jz/U$, which can be determined from the methods introduced here.

\section{Conclusion}

From the discussion in Section \ref{Trap} we conclude that the mean-field approximation yields good results concerning the energy calculated through the one-state approach in Subsection \ref{One-State Approach} as well as by the two-states approach in Subsection \ref{Two-States Approach}. Thus, the particle density (see FIG. \ref{fig:002,008,0101PD}) and the total particle number in a trap (see FIG. \ref{fig.Imu008}) are considered as reliable results. The only physically convincing condensate density stems from the two-states approach (see FIG. \ref{fig:002,008,0101OP} and \ref{20OPS}), whereas the mean-field phase boundary (see FIG. \ref{fig.(1.99)}) is obtained by both the one-state as well as the two-states approach. One way to improve the phase boundary to experimental precision is not to use the mean-field approximation, but a field-theoretic method, where a Legendre transform of the grand-canonical free energy gives very precise results \cite{Ednilson,articleEdnilson}. The same method is supposed to give satisfying results for the superfluid density, which turns out to always coincide with the condensate density in the mean-field picture.

\section{Acknowledgements}

We acknowledge the financial support from the German Research  Foundation within the Collaborative Research Center SBF/TR 49 "Condensed Matter Systems with Variable Many-Body Interactions", SFB/TR 185 "Open System Control of Atomic and Photonic Matter" (OSCAR), and from the binational project DAAD-CAPES. Also, we thank Martin Bonkhoff, Sebastian Eggert, and Carlos S\'a de Melo for helpful discussions. Support from CePOF: 2013/07276-1 is acknowledged. F. T. Sant'Ana acknowledges CAPES for the financial support. F.  E. A. dos Santos acknowledges CNPq for support through Bolsa de produtividade em Pesquisa n.305586/2017-3.

\begin{appendix}

\section{Brillouin-Wigner Perturbation Theory}\label{sec:Appendix A: Derivation of Brillouin-Wigner}

\begin{figure}[t]
	\centering
	\includegraphics[width=.3\textwidth]{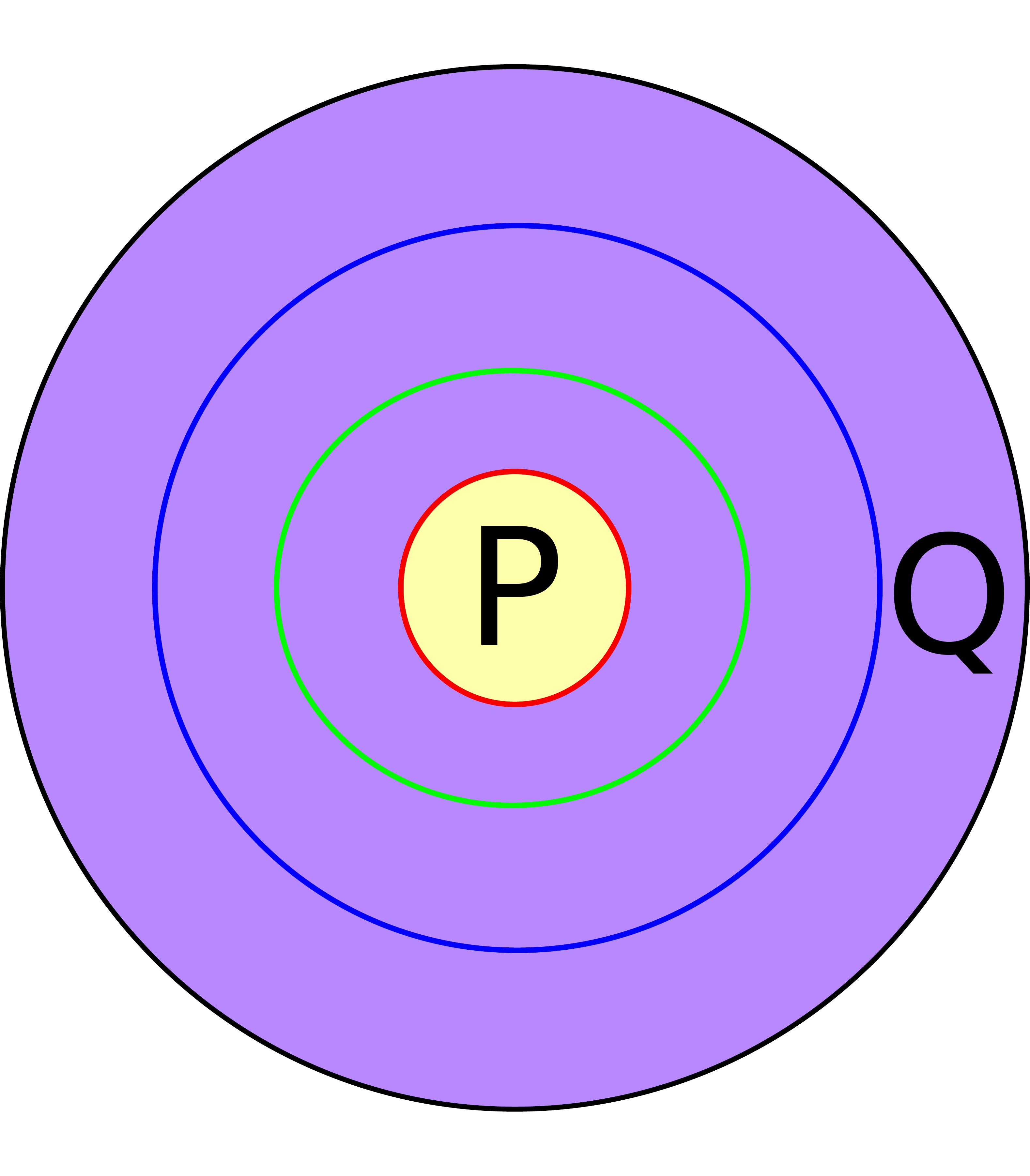}
	\caption{(color online) Generally, the perturbative considerations take place in the infinitely large Hilbert space, which is sketched by the outmost, black ring. In view of a concrete calculation, we have to restrict this space to a finite part, which is illustrated here by the most inner (yellow) circle, labeled by $P$. The infinitely large rest (violet) is labeled by $Q$. The states within $P$ are enough to solve the unperturbed Hamiltonian $\hat{H}^{(0)}$, and can therefore be considered as the zeroth perturbative order with respect to the hopping $J$. For every higher perturbative order, we take more and more of $Q$ into account, just as shown in the figure. Starting from the center, the most inner ring (red) is the zeroth perturbative order, the  second ring (green) stands for the first order, and the third ring (blue) for the second order. For every new perturbative order, a new shell of the $Q$-space encompassing the $P$-space is taken into account, increasing the number of states which are considered.} \label{fig.HilbertSpace}
\end{figure}

Here we provide a concise summary of the Brillouin-Wigner perturbation theory \cite{bookBrillouinWigner}. It amounts to derive an effective Hamiltonian for an arbitrarily chosen Hilbert subspace, which is characterized by a projection operator $\hat{P}$. To this end we have to eliminate the complementary Hilbert subspace, which is characterized by the projection operator $\hat{Q}$, see FIG. \ref{fig.HilbertSpace}.

\subsection{General formalism}
Since we have now two projection operators, i.e. $\hat{P}$ and $\hat{Q}$, we need two conditions to define the respective Hilbert subspaces. So, we start by reformulating the full time-independent Schr\"odinger equation 
\begin{align}\label{eq.2.10,5}
\hat{H}\ket{\Psi_n}=E_{n} \ket{\Psi_n}
\end{align}
with the help of the projection operators. To this end we insert the unity operator $\mathds{1}=\hat{P}+\hat{Q}$ and get
\begin{align}\label{eq.2.13}
\hat{H}\hat{P}\ket{\Psi_n} + \hat{H}\hat{Q}\ket{\Psi_n} = E_{n}\hat{P}\ket{\Psi_n} + E_{n}\hat{Q}\ket{\Psi_n}\,.
\end{align}
Multiplying by $\hat{P}$ the left side of (\ref{eq.2.13}) and considering the projector operator relations $\hat{P}^2=\hat{P}$ and $\hat{P}\hat{Q}=0$ results in
\begin{align}\label{eq.2.14}
\hat{P}\hat{H}\hat{P}\ket{\Psi_n} + \hat{P}\hat{H}\hat{Q}\ket{\Psi_n} = E_{n}\hat{P}\ket{\Psi_n}\,.
\end{align}
Furthermore, multiplying by $\hat{Q}$ the left side of (\ref{eq.2.13}) and using correspondingly $\hat{Q}^2=\hat{Q}$ and $\hat{Q}\hat{P}=0$, we also have
\begin{align}\label{eq.2.15}
\hat{Q}\hat{H}\hat{P}\ket{\Psi_n} + \hat{Q}\hat{H}\hat{Q}\ket{\Psi_n} = E_{n}\hat{Q}\ket{\Psi_n}\,.
\end{align}

The next step is to try to find a single equation for $\hat{P}\ket{\Psi _n}$ in a shape similar to the time-independent Schr\"odinger-equation. In order to eliminate $\hat{Q}\ket{\Psi_n}$ from (\ref{eq.2.14}) we use (\ref{eq.2.15}) and take into account the property $\hat{Q}^2=\hat{Q}$:
\begin{align}\label{eq.2.16}
\hat{Q}\hat{H}\hat{P}\ket{\Psi_n} + \hat{Q}\hat{H}\hat{Q}^2\ket{\Psi_n} = E_{n}\hat{Q}\ket{\Psi_n}\,.
\end{align}
From rearranging and factoring out follows:
\begin{align}\label{eq.2.17}
\hat{Q}\hat{H}\hat{P}\ket{\Psi_n} = \left(E_{n}-\hat{Q}\hat{H}\hat{Q}\right)\hat{Q}\ket{\Psi_n}\,.
\end{align}
Thus, a formal solution with respect to $\hat{Q}\ket{\Psi_n}$ yields
\begin{align}\label{eq.2.18}
\hat{Q}\ket{\Psi_n} = \left(E_{n}-\hat{Q}\hat{H}\hat{Q}\right)^{-1}\hat{Q}\hat{H}\hat{P}\ket{\Psi_n} \,.
\end{align}
A further action of $\hat{Q}$ results in
\begin{align}\label{eq.2.19}
\hat{Q}\ket{\Psi_n} = \hat{Q}\left(E_{n}-\hat{Q}\hat{H}\hat{Q}\right)^{-1}\hat{Q}\hat{H}\hat{P}\ket{\Psi_n}   \,.
\end{align}
Inserting (\ref{eq.2.19}) in (\ref{eq.2.14}), we get a single equation for $\hat{P}\ket{\Psi _n}$:
\begin{align}\label{eq.2.20}
&\left[\hat{P}\hat{H}\hat{P}+\hat{P}\hat{H}
\hat{Q}\left(E_{n}-\hat{Q}\hat{H}\hat{Q}\right)^{-1}
\hat{Q}\hat{H}\hat{P} \right]\ket{\Psi_n} = E_{n}\hat{P}\ket{\Psi_n}\,.
\end{align}
Splitting the Hamiltonian regarding the perturbation allows to rewrite (\ref{eq.2.20}) according to
\begin{align}\label{eq.2.21}
&\hat{P}\hat{H}\hat{P}\ket{\Psi_n}+\hat{P}\left(\hat{H}^{(0)}+\lambda \hat{V} \right)\hat{Q}\left(E_{n}-\hat{Q}\hat{H}\hat{Q}\right)^{-1}\hat{Q}\left(\hat{H}^{(0)}+\lambda \hat{V} \right)\hat{P}\ket{\Psi_n}
= E_{n}\hat{P}\ket{\Psi_n}\,.
\end{align}
From the fact that $\hat{Q}\hat{H}^{(0)}\hat{P}=0$, we finally obtain 
\begin{align}\label{eq.2.23}
&\hat{P} \left[\hat{H}
+\lambda \hat{V} \hat{Q}\left(E_{n}-\hat{Q}\hat{H}\hat{Q}\right)^{-1}\hat{Q}\lambda \hat{V}\right] \hat{P}\ket{\Psi_n}=E_{n}\hat{P}\ket{\Psi_n}\,.
\end{align}
Equation (\ref{eq.2.23}) represents a single equation for $\hat{P} \ket{\Psi_{n}} $, which represents the basis of the Brillouin-Wigner perturbation theory.

\subsection{Matrix representation}
Now we reformulate (\ref{eq.2.23}) in terms of a matrix representation within the Hilbert subspace defined by the projection operator $\hat{P}$. Afterwards, we specialize to the cases that $\hat{P}$ consists of one or two states.

The resulting equation (\ref{eq.2.23}) for $\hat{P}\ket{\Psi _n}$ is of the form of a time-independent Schr\"odinger-equation
\begin{align}\label{eq.2.24}
\hat{P} \hat{H}_{\rm eff}\hat{P}\ket{\Psi_n} = E_{n}\hat{P}\ket{\Psi_n}\,,
\end{align}
where we have introduced the effective Hamiltonian
\begin{align}\label{eq.2.25}
\hat{H}_{\rm eff} = \hat{H}
+\lambda^2 \hat{V} \hat{Q}\left(E_{n}-\hat{Q}\hat{H}\hat{Q}\right)^{-1}\hat{Q} \hat{V}\, .
\end{align}
Since $\hat{H}_{\rm eff}$ is sandwiched by $\hat{P}$ in (\ref{eq.2.24}), everything that goes in or out of $\hat{H}_{\rm eff}$ must involve the Hilbert subspace $\hat{P}$ projects into. However, $\hat{H}_{\rm eff}$ contains also the projection operator $\hat{Q}$, so one has to go beyond the Hilbert subspace $\hat{P}$ projects into.

Another way to represent $\hat{H}_{\rm eff}$ in (\ref{eq.2.25}) is
\begin{align}\label{eq.2.27}
\hat{H}_{\rm eff}=& \hat{H}^{(0)} + \lambda\hat{V}+ \lambda^2 \hat{V} \hat{Q}
  \left(E_{n} - \hat{Q}\hat{H}^{(0)}\hat{Q} - \lambda\hat{Q}\hat{V}\hat{Q} \right)^{-1}\hat{Q} \hat{V}\,.
\end{align}
The resolvent
\begin{align}\label{eq.2.28}
&\hat{R}(E_n)= \left[E_{n} -\hat{Q}\left(\hat{H}^{(0)} + \lambda\hat{V}\right)\hat{Q}\right]^{-1}
\end{align}
can be expanded in series with respect to $\lambda$:
\begin{align}\label{eq.2.29}
\hat{R}(E_n)=&\left(E_{n} -\hat{Q}\hat{H}^{(0)}\hat{Q}\right)^{-1}\sum_{s=0}^{\infty}\left[\lambda\hat{Q}\hat{V}\hat{Q}\left(E_{n} -\hat{Q}\hat{H}^{(0)}\hat{Q}\right)^{-1}\right]^s\,.
\end{align}
Note the crucial property of (\ref{eq.2.29}): instead of the unperturbed energy eigenvalue $E_{n}^{(0)}$ it contains the full energy eigenvalue $E_{n}$.

Inserting (\ref{eq.2.28}) in (\ref{eq.2.27}) results in
\begin{align}\label{eq.2.30,01}
\hat{H}_{\rm eff}=& \hat{H}^{(0)} + \lambda\hat{V}
+ \lambda^2 \hat{V} \hat{Q}\hat{R}(E_n)\hat{Q} \hat{V}\,.
\end{align}
As $\lambda$ approaches zero, this reproduces the unperturbed Schr\"odinger equation. The essential property of (\ref{eq.2.30,01}) is, however, that $E_n$ appears nonlinearly in the resolvent $\hat{R}(E_n)$ from (\ref{eq.2.28}).

Note that the first perturbative order $\lambda\hat{V}$ in (\ref{eq.2.30,01}) is not contained in the resolvent $\hat{R}(E_n)$ but directly emanates from $\hat{H}$. In contrast to that all higher orders in (\ref{eq.2.30,01}) originate from the resolvent term. In particular, $s=0$ gives the second perturbative order, $s=1$ goes up to the third perturbative order and so on. This fundamental difference of origin of perturbative orders is already evident in (\ref{eq.2.13}), where the term $\hat{H}\hat{P}$ gives rise to the zeroth and the first perturbative order, and the term $\hat{H}\hat{Q}$ gives rise to all higher orders. In other words, the zeroth and the first perturbative order are within the Hilbert subspace $\hat{P}$ projects into, whilst for all higher orders, the Hilbert subspace $\hat{Q}$ projects into must be taken into account.

Now we calculate all correction terms of the effective Hamiltonian up to $\lambda ^4$. To do so, we take the sum over $s$ in the resolvent (\ref{eq.2.29}) up to $s=2$ and obtain with (\ref{eq.2.30,01}):
\begin{align}\label{eq.2.30,003}
\nonumber
\hat{H}_{\rm eff}=& \hat{H}^{(0)} + \lambda\hat{V}+ \lambda^2 \hat{V} \hat{Q}\hat{R}^{(0)}(E_n)\hat{Q} \hat{V}+ \lambda^3 \hat{V} \hat{Q}\hat{R}^{(0)}(E_n)
\hat{Q}\hat{V}\hat{Q}\hat{R}^{(0)}(E_n)\hat{Q} \hat{V}
\\
&+ \lambda^4 \hat{V} \hat{Q}\hat{R}^{(0)}(E_n)
\hat{Q}\hat{V}\hat{Q}\hat{R}^{(0)}(E_n)\hat{Q}\hat{V}\hat{Q}\hat{R}^{(0)}(E_n)\hat{Q} \hat{V}\,.
\end{align}
Here we have introduced the resolvent with the unperturbed Hamiltonian
\begin{align}\label{eq.Resolvent}
\hat{R}^{(0)}(E_n) = \left(E_{n} -\hat{Q}\hat{H}^{(0)}\hat{Q}\right)^{-1}\,.
\end{align}
Now we specialize to the respective projection operators $\hat{P}=\sum_{k\in N} \hat{P}_k$ and $\hat{Q}=\sum_{k\in \tilde{N}} \hat{P}_k$, where $\hat{P}_k = \ket{E_k^{(0)}} \bra{E_k^{(0)}}$ represents a projector for the unperturbed eigenstate $\ket{E_k^{(0)}}$. Note that $N$ defines a finite set of quantum numbers, whereas $\tilde{N}$ represents its complement. With this we show that the matrix element of the resolvent (\ref{eq.Resolvent}) yields
\begin{align}\label{eq.2.31,2}
\frac{1}{E_n - E_{l}^{(0)}} = \bra{\Psi_{l}^{(0)}}\hat{R}^{(0)}(E_n)\ket{\Psi_{l}^{(0)}}\,,
\end{align}
with $l \in \tilde{N}$ and $n \in N$. Taking into account (\ref{eq.2.31,2}) in (\ref{eq.2.30,003}), we obtain
\begin{align}\label{eq.2.31,003}
\nonumber
\hat{H}_{\rm eff}=& \hat{H}^{(0)} + \lambda\hat{V}+ \lambda^2 \sum_{l \in \tilde{N}}  \frac{\hat{V}\ket{\Psi_{l}^{(0)}}\bra{\Psi_{l}^{(0)}}\hat{V}}{E_{n}-E_{l}^{(0)}} + \lambda^3 \sum_{l,l' \in \tilde{N}}  \frac{\hat{V}\ket{\Psi_{l}^{(0)}}\bra{\Psi_{l}^{(0)}}\hat{V}\ket{\Psi_{l'}^{(0)}}\bra{\Psi_{l'}^{(0)}}\hat{V}}{\left(E_{n}-E_{l}^{(0)}\right)\left(E_{n}-E_{l'}^{(0)}\right)}
\\
&+ \lambda^4 \sum_{l,l',l'' \in \tilde{N}} \frac{\hat{V}\ket{\Psi_{l}^{(0)}}\bra{\Psi_{l}^{(0)}}\hat{V}\ket{\Psi_{l'}^{(0)}}
\bra{\Psi_{l'}^{(0)}}\hat{V}\ket{\Psi_{l''}^{(0)}}
\bra{\Psi_{l''}^{(0)}}\hat{V}}{\left(E_{n}-E_{l}^{(0)}\right)\left(E_{n}-E_{l'}^{(0)}\right)\left(E_{n}-E_{l''}^{(0)}\right)}+... \,. 
\end{align}

This representation of the effective Hamiltonian $\hat{H}_{\rm eff}$ has no operators anymore in the denominators, and thus can be used as a starting point for further calculations.


Now we determine an equation for the perturbed ground-state energy $E_m$. To this end, we choose $n, n' \in N$ and reformulate (\ref{eq.2.24}) with $\hat{P}=\sum_{k\in N} \hat{P}_k$:
\begin{align}
\sum_{n, n' \in N}\ket{\Psi_{n}^{(0)}}\bra{\Psi_{n}^{(0)}}\hat{H}_{\rm{eff}}\ket{\Psi_{n'}^{(0)}}
\braket{\Psi_{n'}^{(0)}}{\Psi_{m}}= E_{m} \sum_{ n' \in N}\ket{\Psi_{n'}^{(0)}}\braket{\Psi_{n'}^{(0)}}{\Psi_{m}}\,.
\end{align}
Then we multiply the left side by $\bra{\Psi_{n}^{(0)}}$, 
\begin{align}
\sum_{n,n' \in N}\bra{\Psi_{n}^{(0)}}\hat{H}_{\rm{eff}}\ket{\Psi_{n'}^{(0)}}
\braket{\Psi_{n'}^{(0)}}{\Psi_{m}}= E_{m} \sum_{n,n' \in N}\braket{\Psi_{n}^{(0)}}{\Psi_{n'}^{(0)}}\braket{\Psi_{n'}^{(0)}}{\Psi_{m}}\,,
\end{align}
yielding
\begin{align}\label{eq.PsiSum}
\braket{\Psi_{n'}^{(0)}}{\Psi_{m}} \sum_{n,n' \in N} \left(\bra{\Psi_{n}^{(0)}}\hat{H}_{\rm{eff}}\ket{\Psi_{n'}^{(0)}}- E_{m} \delta_{n,n'} \right) = 0\,.
\end{align}
In order to obtain a non-trivial solution $\braket{\Psi^{(0)}_{n'}}{\Psi_{m}} \neq 0$ from (\ref{eq.PsiSum}), we have to demand
\begin{align}\label{eq.2.31,004}
\rm{Det} \left(\bra{\Psi^{(0)}_{\it{n}}}\hat{H}_{\rm{eff}}\ket{\Psi^{(0)}_{\it{n'}}}-E_{\it m}\delta_{\it{n,n'}}\right)=0\, ,
\end{align}
where the determinant in (\ref{eq.2.31,004}) has to be performed with respect to $n, n' \in N$. Note that (\ref{eq.2.31,004}) defines $E_m$ as a zero of a polynomial of finite order.

\subsection{\label{subonestate}Special cases}
Now we specialize (\ref{eq.2.31,004}) to the case that the projector $\hat{P}$ consists of one or two states, respectively.
\subsubsection{One-state approach}
Here we consider first the special case that $\hat{P}$ contains only one state, namely
\begin{align}\label{eq.2.30,04}
\hat{P}=\hat{P}_n\,.
\end{align}
In this case, where $n=n'=m$, (\ref{eq.2.31,004}) simplifies to
\begin{align}\label{eq.2.31,005}
E_n = \bra{\Psi^{(0)}_{n}}\hat{H}_{\rm{eff}}\ket{\Psi^{(0)}_{n}}\,.
\end{align}
Inserting (\ref{eq.2.31,003}) in (\ref{eq.2.31,005}) we get
\begin{align}
\nonumber
E_n &=E^{(0)}_{n} + \lambda V_{n,n}+ \lambda^2 \sum_{l \neq n}  \frac{V_{n,l}V_{l,n}}{E_{n}-E_{l}^{(0)}} + \lambda^3 \sum_{l,l' \neq n}  \frac{V_{n,l}V_{l,l'}V_{l',n}}{\left(E_{n}-E_{l}^{(0)}\right)\left(E_{n}-E_{l'}^{(0)}\right)}
\notag
\\
&+ \lambda^4 \sum_{l,l',l'' \neq n} \frac{V_{n,l}V_{l,l'}V_{l',l''}V_{l'',n}}{\left(E_{n}-E_{l}^{(0)}\right)\left(E_{n}-E_{l'}^{(0)}\right)\left(E_{n}-E_{l''}^{(0)}\right)}+... \,,\label{eq.2.31,006}
\end{align}
where we have taken into account that $\bra{\Psi^{(0)}_n} \hat{H}^{(0)} \ket{\Psi^{(0)}_n} = E^{(0)}_{n}$ and defined the matrix element $V_{n,m} \equiv \bra{\Psi^{(0)}_{n}}\hat{V}\ket{\Psi^{(0)}_{m}}$.

Note that, due to the non-linear appearance of $E_n$, Eq.~(\ref{eq.2.31,006}) represents a self-consistency equation for the energy eigenvalue $E_n$. Furthermore, we observe up to third order that every order in $\lambda$ consists of only one single term. Since we have $n \neq l,l',l''$, the denominator is never zero and thence no divergence occurs in this perturbative representation for the perturbed ground-state energy $E_n$.

\subsubsection{Two-states approach}
Now we consider the case that $\hat{P}$ consists of two states
\begin{align}\label{eq.2.34,9}
\hat{P}=\hat{P}_{n} + \hat{P}_{n'} \,.
\end{align}
Thus, (\ref{eq.2.31,004}) reduces to
\begin{align}\label{eq.2.37}
\rm{Det} \begin{pmatrix}
H_{{\rm eff}, n, n} - E_{m}
& H_{{\rm eff}, n, n'}
\\
H_{{\rm eff}, n', n}
& H_{{\rm eff}, n', n'} - E_{m}
\end{pmatrix}
=
0\,.
\end{align}
Note that
\begin{align}\label{eq.2.38}
\Gamma =
\begin{pmatrix}
H_{{\rm eff}, n, n}
& H_{{\rm eff}, n, n'}
\\
H_{{\rm eff}, n', n}
&H_{{\rm eff}, n', n'}
\end{pmatrix}
\end{align} 
represents a $2 \times 2$   matrix, since  the projection operator $\hat P$ in (\ref{eq.2.34,9}) consists of two states.

\section{Graphical Approach}

In order to evaluate (\ref{eq.2.37}) for higher orders in $\lambda$, it is mandatory to evaluate the matrix elements (\ref{eq.2.38}) from the effective Hamiltonian (\ref{eq.2.31,003}) to higher orders in $\lambda$. To this end we work out here an efficient graphical approach.

\begin{figure}[t]
	\centering
\includegraphics[width=\textwidth,clip]{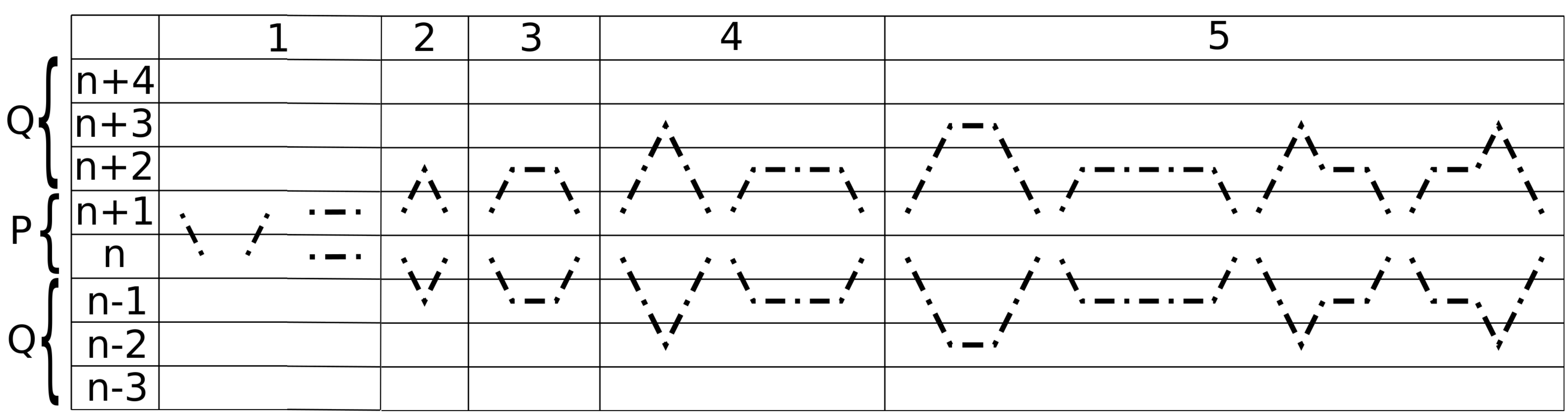}
\caption{Graphical approach for the matrix elements (\ref{eq.2.38}) of the effective Hamiltonian (\ref{eq.2.31,003}) for the Bose-Hubbard mean-field Hamiltonian (\ref{eq.MFHamil}) up to fifth order in the hopping for the two-states approach.\label{figIGA}} 
\end{figure}

In particular, we specify Appendix \ref{sec:Appendix A: Derivation of Brillouin-Wigner} to the mean-field Hamiltonian (\ref{eq.MFHamil}) and find for the two-states approach a graphical representation of the matrix elements in FIG. \ref{figIGA}. The numbers in the first row of FIG. \ref{figIGA} represent the orders of $\lambda$ for the respective correction terms. In the first column we have the different states ranging from $n-3$ to $n+4$. Within the two-states matrix approach we choose $\hat{P}=\hat{P}_n + \hat{P}_{n+1}$, once there is a degeneracy between two consecutive Mott lobes in the zero-temperature phase diagram of the Bose-Hubbard model.

In order to obtain all possible graphs in FIG. \ref{figIGA}, we have to take into account the following empirical rules:
\begin{itemize}
\item{According to $l \in \tilde{N}$ and thus $l \neq n$ in (\ref{eq.2.31,006}), the state we start in and the state we end in can not be reached in between;}
\item{Since $\hat{V}$ is linear in $\hat{a}$ and $\hat{a}^{\dagger}$ in (\ref{V}), we can only get from one state to its nearest neighboring states;}
\item{Because the effective Hamiltonian $\hat{H}_{\rm{eff}}$ in (\ref{eq.2.25}) contains only the projection operator $\hat{Q}$, but is sandwiched by the projection operator $\hat{P}$ according to (\ref{eq.2.24}), it is only allowed that the first and the last state is within $\hat{P}$. This rule actually only occurs for the terms in the diagonal matrix elements.}
\end{itemize}

We interpret each graph according to the following rules:
\begin{itemize}
\item{For every graph we draw the starting point corresponding to
\begin{align}
S \left( \eta \right) = E_n - E_{\eta}^{(0)}\,,
\end{align}
with $\eta$ being the state we start the graph in.}
\item{For every line we draw, we get the following terms. For an ascending line we have
\begin{align}
L_A \left( \nu \right) = - \lambda J z \Psi \frac{\sqrt{\nu +1}}{E_n - E^{(0)}_{\nu}}\,,
\end{align}
with $\nu$ being the state the line started in. For every descending line we draw we get
\begin{align}
L_D \left( \nu \right) = - \lambda J z \Psi^* \frac{\sqrt{\nu}}{E_n - E^{(0)}_{\nu}}\,,
\end{align}
with $\nu$ being the state the line started in.}
\item{For a horizontal line, we get
\begin{align}
L_{H}(\nu)=\frac{\lambda J z \Psi^{*} \Psi}{E_{n}-E_{\nu}^{(0)}}\,,
\end{align}
with $\nu$ being the state the line started in.}
\end{itemize}

In the column labeled as $1$, which corresponds to the order $\lambda$, we have the off-diagonal matrix elements
\begin{align}\label{Gleichung5.18}
S(n+1) L_D(n+1)=-\lambda Jz \Psi^* \sqrt{n+1}\,,
\\
S(n) L_A(n)=-\lambda Jz \Psi \sqrt{n+1}\,,
\end{align}
and the diagonal matrix elements
\begin{align}
S(n+1)L_H(n+1)=\lambda J z \Psi^* \Psi\,,
\\
S(n)L_H(n)=\lambda J z \Psi^* \Psi\,.
\end{align}
For $\lambda^2$ we have correspondingly
\begin{align}
S(n+1) L_A(n+1)L_D(n+2)= \lambda^2 J^2 z^2 \Psi^* \Psi \frac{n+2}{E_n-E_{n+2}^{(0)} }\,
\end{align}
and
\begin{align}
S(n) L_D(n)L_A(n-1)= \lambda^2 J^2 z^2 \Psi^* \Psi \frac{ n}{E_n-E_{n-1}^{(0)} }\,.
\end{align}
For $\lambda^3$ one yields
\begin{align}
S(n+1) L_A(n+1)L_H(n+2)L_D(n+2)= \lambda^3 J^3 z^3 \Psi^{*2} \Psi^2 \frac{n+2}{\left(E_n-E_{n+2}^{(0)}\right)^2 }\,
\end{align}
together with
\begin{align}
S(n) L_D(n)L_H(n-1)L_A(n-1)= \lambda^3 J^3 z^3 \Psi^{*2} \Psi^2 \frac{ n}{\left(E_n-E_{n-1}^{(0)}\right)^2 }\,.
\end{align}
For $\lambda^4$ we find
\begin{align}
\notag
&S(n+1) L_A(n+1)\left[L_A(n+2)L_D(n+3) +  L_H(n+2)L_H(n+2)\right]L_D(n+2)
\\
&= \lambda^4 J^4 z^4 \Psi^{*2} \Psi^2 \frac{ \left(n+2\right) \left(n+3\right)}{\left(E_n-E_{n+2}^{(0)}\right)^2 \left(E_n-E_{n+3}^{(0)}\right)}+ \lambda^4 J^4 z^4 \Psi^{*3} \Psi^3 \frac{n+2}{\left(E_n-E_{n+2}^{(0)}\right)^3 }\,
\end{align}
and
\begin{align}\label{Gleichung5.27}
\notag
&S(n) L_D(n)\left[L_D(n-1)L_A(n-2)+L_H(n-1)L_H(n-1)\right]L_A(n-1)
\\
&= \lambda^4 J^4 z^4 \Psi^{*2} \Psi^2 \frac{ n\left(n-1\right)}{\left(E_n-E_{n-1}^{(0)}\right)^2 \left(E_n-E_{n-2}^{(0)}\right)}+\lambda^4 J^4 z^4 \Psi^{*3} \Psi^3 \frac{n}{\left(E_n-E_{n-1}^{(0)}\right)^3 }\,.
\end{align}
Finally, the fifth column, corresponding to $\lambda^5$, gives:
\begin{align}
\notag
&S(n+1) L_A(n+1)\left[
L_A(n+2)L_H(n+3)L_D(n+3)\right.+L_H(n+2)L_H(n+2)L_H(n+2)
\\
\notag
&+\left.2L_A(n+2)L_D(n+3)L_H(n+2)
\right]L_D(n+2)
\\
\notag
&=\lambda^5 J^5 z^5 \Psi^{*3} \Psi^3 \frac{\left(n+2\right) \left(n+3\right)}{\left(E_n-E_{n+2}^{(0)}\right)^2 \left(E_n-E_{n+3}^{(0)}\right)^2} +2\lambda^5 J^5 z^5 \Psi^{*3} \Psi^3 \frac{\left(n+2\right) \left(n+3\right)}{\left(E_n-E_{n+2}^{(0)}\right)^3 \left(E_n-E_{n+3}^{(0)}\right)}
\\
&+\lambda^5 J^5 z^5 \Psi^{*4} \Psi^4 \frac{n+2}{\left(E_n-E_{n+2}^{(0)}\right)^4 }\,,
\end{align}
together with
\begin{align}
\notag
&S(n) L_D(n)\left[
L_D(n-1)L_H(n-2)L_A(n-2)\right.+L_H(n-1)L_H(n-1)L_H(n-1)
\\
\notag
&+\left.2L_D(n-1)L_A(n-2)L_H(n-1)\right]L_A(n-1)
\\
\notag
&=\lambda^5 J^5 z^5 \Psi^{*3} \Psi^3 \frac{n\left(n-1\right)}{\left(E_n-E_{n-1}^{(0)}\right)^2 \left(E_n-E_{n-2}^{(0)}\right)^2}+\lambda^5 J^5 z^5 \Psi^{*3} \Psi^3 \frac{n\left(n-1\right)}{\left(E_n-E_{n-1}^{(0)}\right)^3 \left(E_n-E_{n-2}^{(0)}\right)}
\\
&+\lambda^5 J^5 z^5 \Psi^{*4} \Psi^4 \frac{n}{\left(E_n-E_{n-1}^{(0)}\right)^4} \,.
\end{align}

\section{\label{appC}Superfluid Density for Mean-Field}

The mean-field Hamiltonian (\ref{H0})-(\ref{H}) is local and has the form
\begin{align}
\hat{H}=h\left( \hat{n} \right) + j \hat{a} + j^{*} \hat{a}^{\dagger}\,,
\end{align}
where $h\left( \hat{n} \right)$ stands for the local term $Jz\Psi^*\Psi+U\hat{n}(\hat{n-1})/2 -\mu \hat{n}$, while the currents correspond to $j=-Jz\Psi$ and $j^*=-Jz\Psi^*$. Its ground-state energy is
\begin{align}
E_0={\cal{E}}\left( j^{*} j \right)\,,
\end{align}
and the energy will then be 
\begin{align}\label{meanfieldfreeeergy}
E=N_{s}\left[ Jz |\psi|^2 + {\cal{E}} \left( J^2 z^2 |\psi|^2 \right) \right] \,.
\end{align}
Considering a Galilei boost $z \rightarrow  z - \left( \frac{a}{L} \vec{\phi} \right)^2$ results in
\begin{align}\label{APPCGALI}
E\left[ \varepsilon \left(\vec{\phi} \right) \right] = N_s  J \varepsilon \left( \vec{\phi} \right) \rho_c \left[ \varepsilon \left( \vec{\phi} \right) \right] + N_s {\cal{E}}\left( \left[ J \varepsilon \left( \vec{\phi} \right) \right]^2 \rho_c \left[ \varepsilon \left( \vec{\phi} \right) \right] \right) \,,
\end{align}
with
\begin{align}
\varepsilon \left( \vec{\phi} \right)=2\sum_{l}{\rm{cos}}\left( \frac{a}{L} \phi_l \right)\,,
\end{align}
where $\rho_c \left( \vec{\phi} \right)$ is the $\phi$-dependent condensate density satisfying the equation
\begin{align}\label{APPCOBEY}
J \varepsilon \left( \vec{\phi} \right)+ \left[J \varepsilon \left(\vec{ \phi} \right)\right]^2 {\cal{E}}'\left( \left[J \varepsilon \left( \vec{\phi} \right)\right]^2 \rho_c  \left( \vec{\phi} \right) \right) =0\,.
\end{align}
Therefore, the superfluid density is given by \cite{Bradlyn}
\begin{align}
\rho_{\rm{SF}}=\lim_{\vec{\phi} \to \vec{0}}\frac{L^2 \left\{ E\left[ \varepsilon \left( \vec{\phi} \right) \right] - E\left( z \right) \right\}}{Ja^2 N_s \sum_{l}\phi^{2}_{l} } \,,
\end{align}
resulting in
\begin{align}
\rho_{\rm{SF}}=-\frac{E^\prime\left( z \right)}{J N_s}\,.
\end{align}
On the other hand, differentiating (\ref{APPCGALI}) yields
\begin{align}
E'\left( z \right)=N_s \left[ J \rho_c \left( z \right)+Jz \rho_{c}' \left( z \right)-\left( 2J^2 z \rho_c \left( z \right) +J^2 z^2 \rho_{c}' \left( z \right) \right){\cal{E}}' \left( J^2 z^2 \rho_c \left( z \right) \right) \right]\,.
\end{align}
By using (\ref{APPCOBEY}) we get
\begin{align}
E'\left( z \right) = - N_s J \rho_c \left( z \right)\,,
\end{align}
and therefore
\begin{align}
\rho_{\rm{SF}}=\rho_c \left( z \right)\,.
\end{align}
Thus we conclude that superfluid and condensate density must always be equal for the mean-field theory.

\end{appendix}

\bibliographystyle{apsrev}

\end{document}